\documentclass[apj]{emulateapj}

\usepackage{color}
\usepackage{natbib}

\newcommand\ionB[2]{#1$\;${\scshape{#2}}}

\def\ForbSII{\mbox{[\ionB{S}{ii}]}}

\def\MgII{\mbox{\ionB{Mg}{ii}}}

\def\Ha{\ifmmode^{\mathrm{H}\alpha }\else$\mathrm{H}\alpha$\fi}
\def\Hb{\ifmmode^{\mathrm{H}\beta }\else$\mathrm{H}\beta$\fi}
\def\LyA{\ifmmode^{\mathrm{H}\alpha }\else$\mathrm{Ly}\alpha$\fi}
\def\BrA{\ifmmode^{\mathrm{Br}\alpha }\else$\mathrm{Br}\alpha$\fi}
\def\BrG{\ifmmode^{\mathrm{Br}\gamma }\else$\mathrm{Br}\gamma$\fi}
\def\PaB{\ifmmode^{\mathrm{Pa}\beta }\else$\mathrm{Pa}\beta$\fi}
\def\mag{\ifmmode^{\rm m }\else$^{\rm m}$\fi}

\def\as{$\,^{\prime\prime}\,$}

\def\hh{\ifmmode^{\rm h}\else$^{\rm h}$\fi}
\def\mm{\ifmmode^{\rm m}\else$^{\rm m}$\fi}
\def\ss{\ifmmode^{\rm s}\else$^{\rm s}$\fi}
\def\deg{\ifmmode^\circ\else$^\circ $\fi}
\def\amin{\ifmmode^\prime\else$^\prime $\fi}

\def\decdm#1#2{\ifmmode{#1}\else{$#1$}\fi\deg\ #2\amin\ }

\def\dec#1#2#3{\ifmmode{#1}\else{$#1$}\fi\deg\ #2\amin\ #3\as\ }

\def\decb#1#2#3#4{\ifmmode{#1}\else{$#1$}\fi\deg\ #2\amin\ #3\farcs#4 }

\shorttitle{V921\,Scorpii: Geometry and kinematics of the circumprimary disk on sub-AU scales}
\shortauthors{Kraus et al.}

\begin{document}


\title{On the nature of the Herbig~B[e] star binary system V921\,Scorpii:\\
  Geometry and kinematics of the circumprimary disk on sub-AU scales\footnotemark[1]}

\footnotetext[1]{Based on observations made with ESO telescopes 
at the Paranal Observatory under the open-time program ID 
084.C-0668(A, B) and with the Magellan Clay telescope.
}


\author{Stefan Kraus\altaffilmark{1}, 
  Nuria Calvet\altaffilmark{1},
  Lee Hartmann\altaffilmark{1},
  Karl-Heinz Hofmann\altaffilmark{2},
  Alexander Kreplin\altaffilmark{2},
  John D.\ Monnier\altaffilmark{1}, and
  Gerd Weigelt\altaffilmark{2}}


\affil{
$^{1}$~Department of Astronomy, University of Michigan, 918 Dennison Building, Ann Arbor, MI 48109-1090, USA\\
$^{2}$~Max Planck Institut f\"ur Radioastronomie, Auf dem H\"ugel 69, 53121 Bonn, Germany
}


\begin{abstract}
V921\,Scorpii is a close binary system (separation {0.025\arcsec})
showing the B[e]-phenomenon.  The system is surrounded by an enigmatic 
bipolar nebula, which might have been shaped by episodic mass-loss events,
possibly triggered by dynamical interactions between the companion 
and the circumprimary disk \citep{kra12b}.
In this paper, we investigate the spatial structure and kinematics of the 
circumprimary disk, with the aim to obtain new insights into the still
strongly debated evolutionary stage.
For this purpose, we combine, for the first time, 
infrared spectro-interferometry (VLTI/AMBER, $\lambda/\Delta\lambda=12,000$) and spectro-astrometry 
(VLT/CRIRES, $\lambda/\Delta\lambda=100,000$),
which allows us to study the AU-scale distribution of circumstellar gas 
and dust with an unprecedented velocity resolution of 3~km\,s$^{-1}$.
Using a model-independent photocenter analysis technique, we find that the {\BrG}-line emitting gas
rotates in the same plane as the dust disk.
We can reproduce the wavelength-differential visibilities and phases and the double-peaked 
line profile using a Keplerian-rotating disk model. 
The derived mass of the central star is $5.4\pm 0.4~M_{\sun} \cdot (d/1150\mathrm{pc})$,
which is considerably lower than expected from the spectral classification,
suggesting that V921\,Sco might be more distant ($d \sim 2$~kpc) 
than commonly assumed.
Using the geometric information provided by our {\BrG} spectro-interferometric data
and Paschen, Brackett, and Pfund line decrement measurements
in 61 hydrogen recombination line transitions, 
we derive the density of the line-emitting gas ($N_{e}=2...6\times 10^{19}$\,m$^{-3}$).
Given that our measurements can be reproduced with a Keplerian velocity field 
without outflowing velocity component 
and the non-detection of age-indicating spectroscopic diagnostics, 
our study provides new evidence for the pre-main-sequence nature of V921\,Sco.
\end{abstract}





\section{Introduction}

The class of the B[e] stars comprises some of the most 
peculiar and mysterious objects of stellar astrophysics.  
In particular, it has been found that the class-defining 
characteristics, which include infrared excess and
strong permitted and forbidden line emission,
are observed in a wide range of evolutionary stages,
including stars of pre-main-sequence, 
post-main-sequence, and unknown nature 
(unclassified B[e] stars, \citealt{lam98}).
An enigmatic member of the latter category is \object{V921\,Scorpii}
(=Hen\,3-1300, MWC\,865, CD\,-42$^{\circ}$11721).
Despite more than one hundred studies that have been
conducted over the last four decades, 
the stellar parameters and evolutionary stage of V921\,Sco 
remain strongly debated.
These uncertainties arise primarily from problems in
distance estimation, where values between 
160~pc \citep{hil92} and 2.6~kpc \citep{sho90}
have been proposed.
Furthermore, due to the lack of photospheric lines in the optical spectrum,
the existing spectral classifications are based on 
rather indirect methods, such as photometric modeling 
or the ionization analysis of circumstellar emission lines.
Accordingly, the derived effective temperatures 
(12,300~K to 31,600~K; \citealt{hil92,cid01}) and spectral types
(Aep to B0[e]p; \citealt{dew90,nat93}) exhibit a
considerable spread, which prevents a clear 
Hertzsprung-Russell diagram classification and has lead authors 
to argue both for an evolved (supergiant; \citealt{hut90,bor07}) and
young (Herbig~B[e]; \citealt{dew90,the94,ben98,hab03,ack05,ack06}) 
object nature.

The aim of this paper is to gain new insights on the astrophysical
nature of V921\,Sco, and B[e] stars in general, by studying the 
spatial distribution and kinematics of the circumstellar gas and dust.  
These constraints will inform us about the disk-formation mechanism 
(i.e.\ accretion versus excretion), and, thus, the evolutionary state of V921\,Sco.
For instance, accretion disks around young stars are believed to exhibit a
Keplerian rotation profile, while the radiation-driven winds from
evolved stars should exhibit a strong outflowing velocity component \citep{lam91}.
Given the possible kiloparsec distance of V921\,Sco, this task requires 
both high angular and high spectral resolution,
which we achieve in our study by combining, for the first time,
two highly complementary techniques, namely spectro-interferometry 
(VLTI/AMBER, providing milliarcsecond angular resolution and spectral resolution $R=12,000$)
and spectro-astrometry (VLT/CRIRES, $R=100,000$).
From these spatially and spectrally resolved constraints, we
derive the gas velocity field on scales down to a few stellar radii.

In a recent study \citep[][Paper~I]{kra12b}, we already obtained milliarcsecond-resolution
continuum interferometric images of V921\,Sco in three wavelength bands (1.65, 2.0, and 2.3~$\mu$m)
and discovered a close (separation $\rho \sim 25$~milliarcseconds=mas) companion, 
as well as a continuum-emitting disk with an apparent size of $7.5 \pm 0.2$~mas 
(Gaussian FWHM along major axis).
In addition, we obtained images of the surrounding large-scale bipolar nebula
and detected multi-layered, shell-like substructures that
might have been shaped by episodic mass-loss events. 
Based on roughly matching timescales between the estimated orbital period and the 
mass-ejection period, we suggested that the mass-ejection events 
might be triggered by the newly discovered companion.

For the stellar parameters of V921\,Sco, we adopt in this study 
a luminosity $L_{\star}=(10\pm3) \times 10^{3} L_{\sun}$, 
effective temperature $T_{\rm eff}=(14\pm1) \times 10^{3}$~K,
a distance of $d=1150\pm150$~pc, $A_V=4.8 \pm 0.2$, 
and a stellar radius of $R_{\star} = 17.3 \pm 0.6~R_{\sun}$ \citep{bor07}.
The implications of a modified set of fundamental parameters
will be discussed in Sect.~\ref{sec:interpdisk}.

In the following, we present our spectro-interferometric (Sect.~\ref{sec:obsAMBER}),
spectro-astrometric (Sect.~\ref{sec:obsCRIRES}), and 
spectroscopic (Sect.~\ref{sec:obsFIRE}) observations, followed by a discussion and 
quantitative modeling, both in the continuum (Sect.~\ref{sec:continuum}) and 
the {\BrG}-line (Sect.~\ref{sec:line}).
Finally, we will interpret our modeling results (Sect.~\ref{sec:interp}) and
conclude with a brief summary of our findings (Sect.~\ref{sec:conclusions}).

\section{Observations}
\label{sec:observations}

\subsection{VLTI/AMBER spectro-interferometry}
\label{sec:obsAMBER}

\begin{figure}
  \centering
  \includegraphics[angle=0,scale=1.]{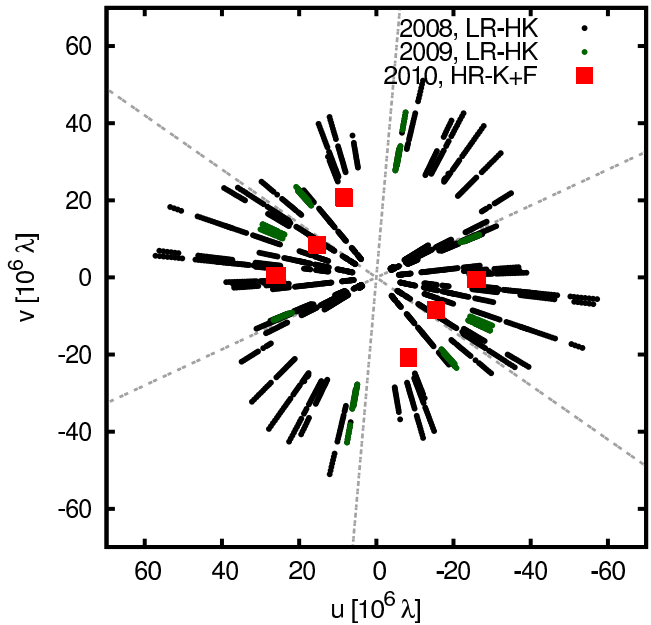}
  \caption{
    $uv$-coverage of our VLTI/AMBER observations using the LR-HK mode ({\it black dots}, Paper~I) and
    HR-K mode ({\it red squared}, this paper), as well as the
    slit orientation of the spectro-astrometric observations ({\it dashed grey lines}).
  }
  \label{fig:uvcov}
\end{figure}

\begin{figure*}
  \centering
  \includegraphics[angle=0,scale=0.8]{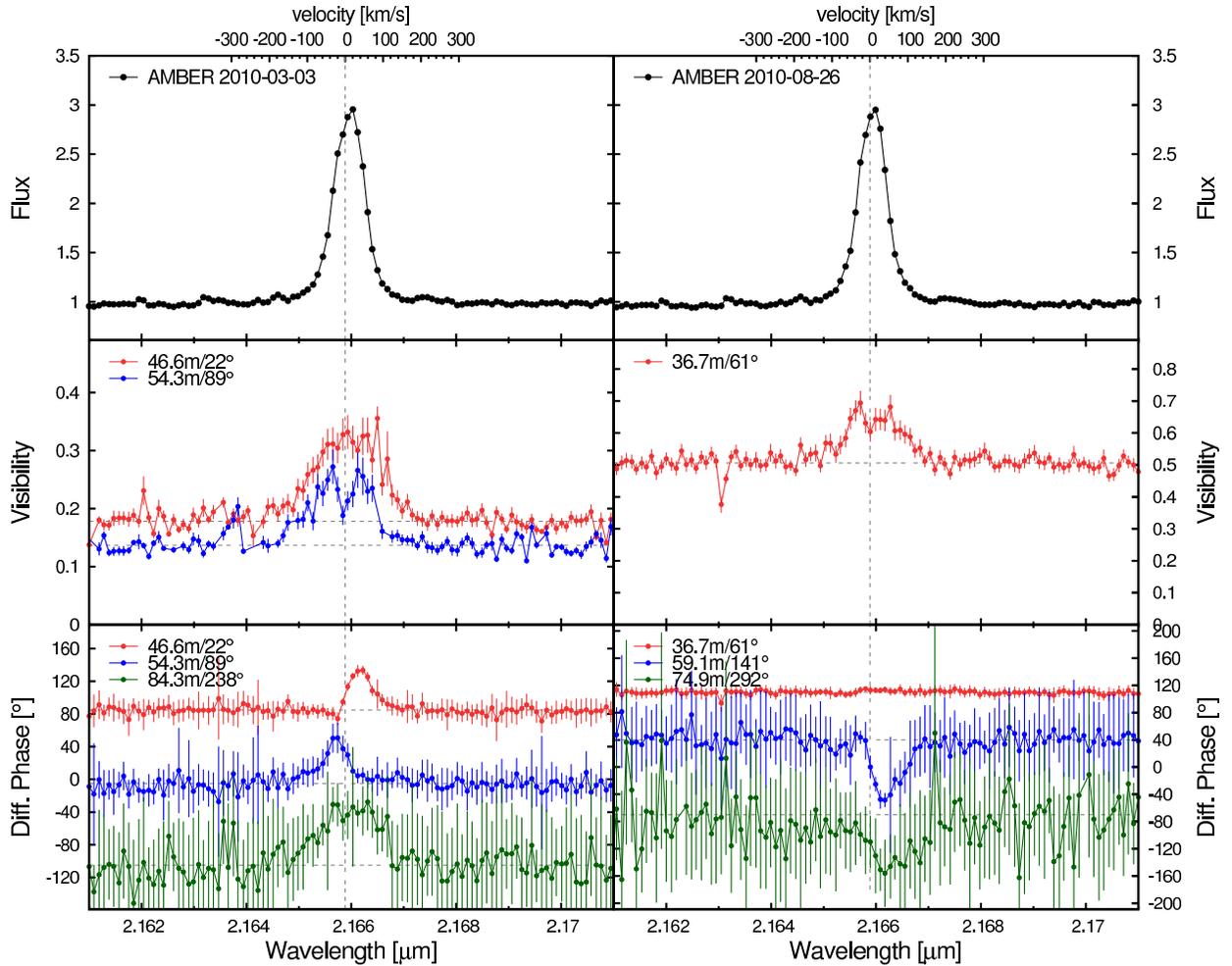} \\
  \caption{
    Spectra {\it (top)}, visibilities {\it (middle)}, and DPs {\it (bottom)} derived
    from our VLTI/AMBER observations with spectral resolution $R=12,000$ obtained on 
    2010-03-03 {\it (left)} and 2010-08-26 {\it (right)}.
    For the DP measurements, we add an arbitrary offset (dashed line) for clarity.
  }
  \label{fig:AMBERoverview}
\end{figure*}

Infrared interferometers, such as ESO's Very Large Telescope Interferometer (VLTI),
coherently combine the light from separate apertures in order to 
achieve an unprecedented angular resolution of a few milliarcseconds
at infrared wavelengths.
With a spectral resolving power of up to $R=\lambda/\Delta\lambda=12,000$, these instruments now enable 
investigations both in the continuum emission and in spatially and spectrally resolved gas-tracing lines.
The measured visibility amplitudes, closure phases, and wavelength-differential phases (DPs)
provide powerful constraints for model-fitting or can be used for the reconstruction of interferometric images.

First spectro-interferometric observations on V921\,Sco 
using the VLTI near-infrared beam combiner instrument AMBER \citep{pet07}
were presented by \citet{kra08b}, \citet{kre12}, and in Paper~I
and provided low (LR-HK, $R=35$) and medium (MR-K, $R=1500$) spectral resolution.
Here, we present new observations using AMBER's unique high spectral resolution mode 
(HR-K, $R=12,000$).
The new data sets were obtained on 2010-03-03 and 2010-08-26 
using the 8.2\,m unit telescopes UT2, UT3, and UT4
and cover a wavelength window around the {\BrG} 2.16602~$\mu$m line.

In order to yield sufficient signal-to-noise ratio (SNR) for our high spectral resolution observations,
we employed the FINITO fringe tracker instrument \citep{gai04,leb08}.
This instrument measures and corrects the atmosphere-induced phase perturbations
and allowed us to record the optical path delay (OPD)-stabilized interferograms with a long
detector integration time of 3\,s.
Unfortunately, the tracking performance of FINITO was
degraded for the longest baselines due to the
low visibility contrast in this baseline regime.
Based on the poor SNR, we rejected the visibility measurements
for one baseline from 2010-03-03 (UT2-UT4)
and two baselines from 2010-08-26 (UT2-UT4, UT3-UT4).
The remaining baselines exhibit fringe SNRs up to
3.5, 4.5, and 14, and cover baseline lengths between 
36.7 and 84.3~m and PAs between 22 and $292^{\circ}$ (Fig.~\ref{fig:uvcov}).
The data shows a double-peaked, rising visibility profile 
in the {\BrG}-line as well as non-zero DPs (Fig.~\ref{fig:AMBERoverview}).

The wavelength calibration was done using atmospheric 
telluric features close to the {\BrG}-line and by
applying a heliocentric-barycentric system correction with
heliocentric velocities of $+28.33$~km\,s$^{-1}$ (2010-03-03) 
and $-27.09$~km\,$^{-1}$ (2010-08-26), respectively.

Each observation on V921\,Sco was accompanied by
observations on the interferometric calibrator \object{HD\,161068},
for which we assume a uniform-disk diameter of $1.448 \pm 0.019$~mas \citep{mer06}.
Both for the science and calibrator star observations, 
we extract raw visibilities, closure phases, and DPs
using the amdlib (V3.0) data reduction software \citep{tat07b,che09}.

Due to the influence of residual vibration, it is known that AMBER+FINITO 
observations with long DIT, such as used for our HR-K mode observations,
might result in a poor absolute visiblity calibration,
while the wavelength-differential observables are only marginally affected \citep[e.g.][]{tat07a}.
To recalibrate the absolute continuum visibility level of the HR-K observations, 
we make use of our earlier LR-HK observations (Paper~I) and the best-fit continuum
model (Model DISK), which will be described in more detail in Sect.~\ref{sec:continuum}.
We investigated the influence of frame
selection in the course of our data reduction procedure
and find that the wavelength-differential signatures 
are very robust when selecting, for instance, the 
80\%, 50\%, 20\%, or 10\% of frames with the best SNR.
Given that the statistical noise increases with decreasing
frame number and that low-SNR frames are down-weighted 
during the averaging process in a natural fashion, 
we decided to employ the observables from the full, 
unselected data sets for our analysis.

In order to associate the closure phase sign with the
on-sky orientation, we use a reference 
data set\footnote{The reference data set can
be accessed on the website http://www.stefan-kraus.com/files/amber.htm}
on the binary star $\theta^1$\,Orionis~C \citep{kra09a}.
To calibrate the wavelength-differential phases, we make use of
the fact that the continuum photocenter is systematically offset
with respect to the {\BrG}-line emission due to the presence 
of a close companion star.  Using the well-established continuum
closure phase sign calibration, we first determine the 
position angle\footnote{In this paper, all position angles are measured east of north.} (PA)
of the companion star (Sect.~\ref{sec:continuum}) and then adjust the
DP sign in our spectro-interferometric and spectro-astrometric data
in order to match the direction of the continuum photocenter displacement.
This procedure calibrates the DP sign unambiguously,
and might serve also as a reference for other studies using AMBER's HR-mode.
Recently, we employed this calibration in order to determine the
rotation sense of the disk around the classical Be star $\beta$\,CMi \citep{kra12a}.
Since $\beta$\,CMi was observed using both AMBER's MR and HR-mode,
we can also extend the calibration to AMBER MR-mode observations,
which resulted in a recalibration of the $\zeta$\,Tau rotation sense \citep{kra12a}
compared to earlier studies \citep{ste09}.

The statistical error bars on the DPs vary strongly for the different baselines,
which is both a result of the low fringe contrast at the longest baselines
and of differences in the vibration properties of the UTs.
These differences might give overproportional weight to the short baselines 
in our $\chi^2$-fitting procesure.  Therefore, based on the typical
root-mean-square noise in the continuum channels, we include 
a minimum DP error of $20^{\circ}$ for all baselines in the fitting procedure.

\subsection{VLT/CRIRES spectro-astrometry}
\label{sec:obsCRIRES}

\begin{figure}
  \centering
    \includegraphics[angle=270,scale=0.33]{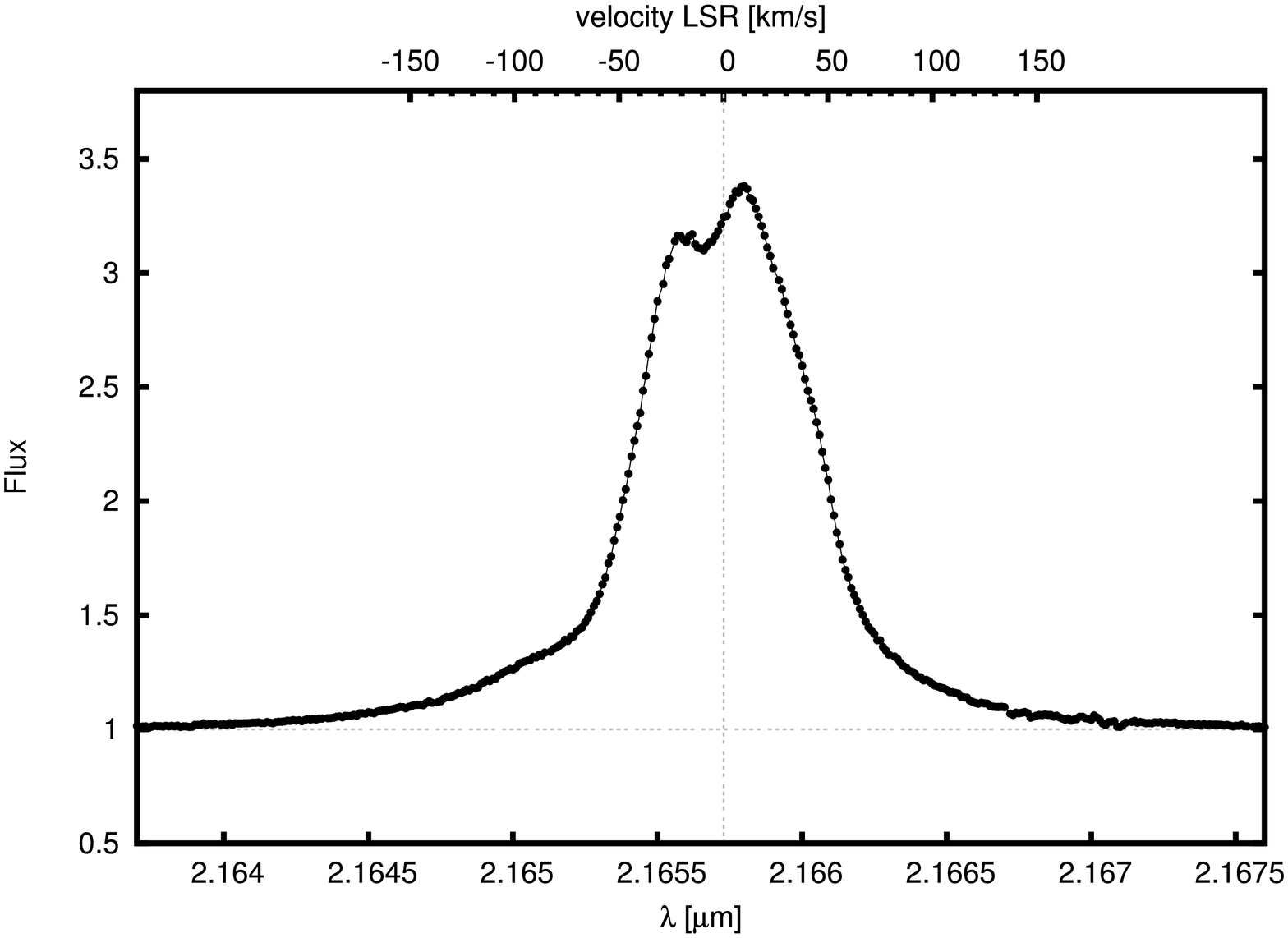} \\
    \includegraphics[angle=270,scale=0.33]{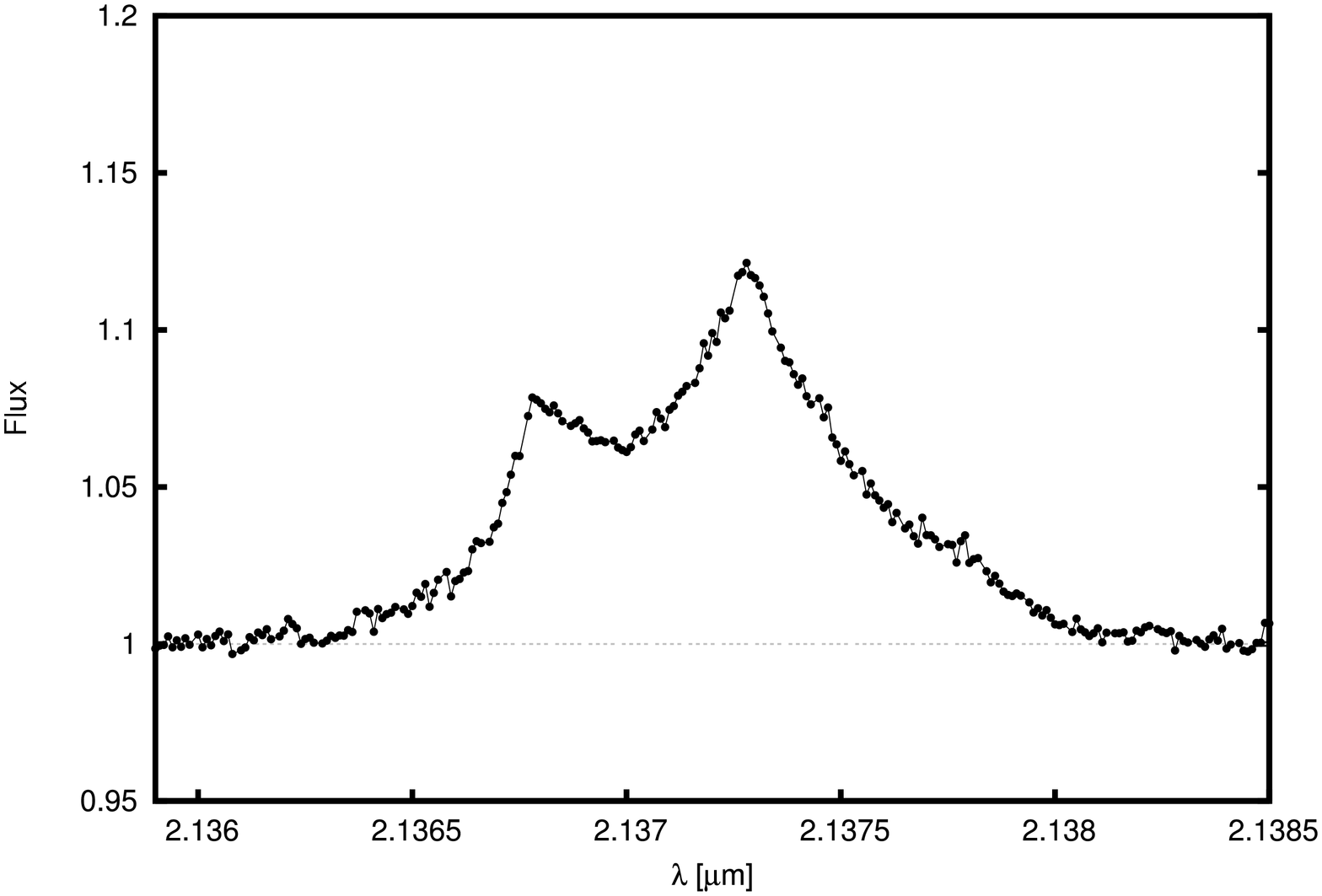} \\
  \caption{
    VLT/CRIRES ($R=100,000$) spectra obtained on V921\,Sco
    in the {\BrG} and {\MgII} line.
  }
  \label{fig:CRIRES}
\end{figure}

Spectro-astrometry uses high signal-to-noise (SNR) long-slit spectra 
to measure the centroid position of an unresolved object as function of wavelength. 
Since the centroid position can be measured with much higher precision than the size of
the point-spread function \citep{bai98a}, this method allows one to measure sub-mas photocenter displacements
in spectrally resolved emission lines. 
While the earliest astronomical applications in the field of star formation focused 
on the detection of close companions \citep[e.g.][]{bai98b}, later studies also 
successfully applied spectro-astrometry to more general cases, such as 
outflow signatures \citep{tak03,whe05}
or the characterization of the gas kinematics in protoplanetary disks 
\citep{pon11,got12}.

Our spectro-astrometric observations on V921\,Sco were obtained using the 
VLT near-infrared high-resolution spectrograph CRIRES \citep{kae04}.
The measurements were obtained on 2010-03-04, UT~07:51 to 08:58
using an integration time of 5\,s and a slit width of 0.2\arcsec, 
resulting in a spectral resolution of $R=100,000$.
In order to apply the spectro-astrometry technique, we recorded spectra
towards three different position angles
({55\deg}, {115\deg}, {175\deg}), and the corresponding
anti-parallel position angles ({235\deg}, {295\deg}, {355\deg}),
complementing the position angles probed by interferometry (Fig.~\ref{fig:uvcov}).
To maximize the flux in the slit, the spectra were recorded using the
STRAP adaptive optics system, resulting in a typical PSF Gaussian FWHM of $\sim 115$~mas.
The measured spectra were corrected to the heliocentric-barycentric system using a
heliocentric velocity of $+28.36$~km\,s$^{-1}$.
The spectra were extracted using the ESO CRIRES data reduction pipeline (version 1.12.0).

Besides the {\BrG}~2.166078~$\mu$m line, which was covered by CRIRES detector \#3,
our observation also covered the {\MgII}~2.137~$\mu$m line in detector \#1.
This line has also been detected in other B[e] stars \citep{cla99}.
To compute velocities in the local standard of rest (LSR), we assumed 
a systemic velocity of $v_{0}=-20$~km\,s$^{-1}$, 
which we adopt from the forbidden line measurements by \citet{bor07}.

In order to correct the spectrum for telluric spectral features, we
observed the early G-type star Hip84425 and  we modeled the 
intrinsic {\BrG}-absorption using an infrared solar spectrum 
recorded with ACE-FTS \citep{has10}.

The final spectra are shown in Fig.~\ref{fig:CRIRES}
and reveal a very narrow (FWHM 0.0006560~$\mu$m or $\approx 91$~km\,s$^{-1}$)
and double-peaked line profile with a very small peak separation of just 
$\sim 0.000184~\mu$m ($\sim 25$~km\,s$^{-1}$).
In addition to the {\BrG}-line spectrum, 
which will be discussed and modeled in Sect.~\ref{sec:line},
we have also recorded the {\MgII}~2.137~$\mu$m line.
For the {\MgII}-line, the peak separation
is significantly wider ($\sim 0.000500~\mu$m or $70$~km\,s$^{-1}$),
indicating that this line originates at smaller
stellocentric radii in the circumstellar environment.
This finding is consistent with our non-detection of a
spectro-astrometric signal within this line.

To derive the astrometric signal, we compute for each spectral channel $v$
the beam centroid position \citep{pon11} in spatial direction
\begin{equation}
  X^{p,a}(v) = K \frac{\sum_{i} x_{i}(v) F_{i}(v)}{\sum_{i} F_{i}(v)},
\end{equation}
where $i$ defines the size of the virtual aperture
(we use $\pm 1$~pixel around the center of the PSF)
and $K$ is defined as the ratio between the flux in the
virtual aperture and the total PSF ($\sim 1.5$).
The astrometric signals are derived for each position angle ($X^{p}$) 
and its anti-parallel counterpart ($X^{a}$) and then subtracted
in order to remove potential artefacts \citep{bra06}:
\begin{equation}
  X(v)=(X^{p}(v)-X^{a}(v))/2.
\end{equation}

The interpretation of spectro-astrometric signals is complicated by the fact that this
technique measures only the first-order momentum in the brightness distribution (relative position),
but is insensitive to higher-order momenta \citep[angular size, asymmetry, kurtosis; see][]{lac03}
as well as the continuum-emitting geometry, which limits this technique to relatively simple cases or requires 
additional model assumptions.
In this study, we make use of the fact that photocenter displacements measured with spectro-astrometry 
are mathematically equivalent to the DPs measured in spectro-interferometry,
which allows us to directly combine the spectro-interferometric (Visibility, DP, closure phases) and 
spectro-astrometric observables (photocenter displacements) for quantitative modeling (Sect.~\ref{sec:linemodeldisk}), 
providing unique constraints on the AU-scale spatial distribution and kinematics of the circumstellar gas and dust.
For this purpose, we translate the measured astrometric signal into the equivalent DP using the relation
\begin{equation}
  \phi = -\frac{2\pi X(v)}{\sigma},
\end{equation}
where $\sigma$ is the FWHM of the PSF measured in the spectrum (typically $\sim 150$~mas).
In order to give a similar weight to the AMBER and CRIRES measurements,
we assume a minimum DP error of $1^{\circ}$ for our CRIRES DPs.

\subsection{Magellan/FIRE near-infared spectroscopy}
\label{sec:obsFIRE}

We recorded a high-resolution, high-SNR near-infrared spectrum of V921\,Sco
using the FIRE Echelle spectrograph \citep{sim08} mounted at the Magellan/Baade 6.5\,m telescope.
With a {0.45\arcsec} slit, this instrument provides a spectral resolution
$R=8000$ and a wide wavelength coverage from 0.8 to 2.5~$\mu$m.
The spectrum was recorded on 2011-03-12 using an A-B-B-A dithering pattern with
DITs of 10\,s (high gain mode, to obtain high SNR in the $z-$, and $J$-band) 
and 5\,s (low gain mode, in order to avoid saturation in the $H$- and $K$-band).
The slit was oriented along the disk polar axis derived from our interferometric observations
(PA=56\deg).
To correct for telluric features, we observed the A0V-type standard star \object{HD\,122945}.
The data reduction was performed using the standard FIRE data reduction pipeline
developed at MIT, which performs also a spectro-photometric calibration based
using the photometry of the standard star.
Trying different combinations of target star / standard star data sets, we find that 
the derived spectral slopes are consistent on the level of a few percent, while the
derived absolute flux levels show some larger scatter in the $\sim 5-10$\% range.
In order to improve on the calibration, we fit a fifth-order polynomical in order to
extract the spectral slope and then use archival ISO photometry 
in order to recalibrate the absolute flux at $2.4\mu$m
which yields also a satisfying match with 2MASS $J$-, $H$-, and $K$-band photometry
(the derived flux densities match within 20\%).
The derived absolute-calibrated FIRE spectrum is corrected for
reddening \citep[$A_V=4.8 \pm 0.2$, ][]{bor07} using the
extinction law from \citet[][$R_V=3.1$]{car89}.

At the longest wavelengths, the SNR in our final spectrum is reduced 
compared to the $J$- and $H$-band since some of the recorded frames had to be rejected due 
to saturation in this part of the spectrum.  In particular, the saturation occurs 
in sky lines, resulting in some narrow spikes 
in the final corrected spectrum.

\section{Results: Continuum geometry}
\label{sec:continuum}

In the following, we investigate the continuum structure of
the circumprimary disks using the VLTI/AMBER low spectral
dispersion data presented in Paper~I.
In this earlier study, the circumprimary disk emission was approximated
with a Gaussian brightness distribution (in the following denoted ``GAUSS'' model), 
which provided a sufficient representation 
to extract the astrometric information for the companion star.
However, the relatively large $\chi^2_r$-value of 4.30 already indicates that this
simple geometry does not provide an appropriate representation of the circumprimary disk structure.
Therefore, we consider here a more realistic parameterization
for the structure of the circumprimary disk, using
a disk temperature-gradient model (in the following denoted ``DISK'' model).

In the DISK model, the emission extends from an inner truncation radius 
$R_{\rm in}$ to an outer radius $R_{\rm out}$ and radiates as a black-body
with $T(r) = T_{\rm in} (r/R_{\rm in})^{-q}$,
where $T_{\rm in}$ denotes the temperature at the inner disk radius and
$q$ is the temperature power-law index.
In addition, we introduce the inclination angle $i$ and position angle $\varphi$,
which define the projection ratio 
($\cos i = R_{\rm minor}/R_{\rm major}$) and on-sky orientation of the disk major axis.
The inclination is measured from the polar axis (i.e.\ $i=0^{\circ}$ is pole-on).
As in our earlier modeling attempts, the DISK model
includes the photospheric emission of the primary and secondary star, 
which are parameterized by the companion separation
($\rho$) and PA ($\Theta$), the angular extension of the circumsecondary material
(given by a Gaussian with FWHM $\theta_{B}$), 
and two parameters ($\frac{F_{\rm B}}{F_{\rm tot}}(2~\mu\mathrm{m})$, $s$) 
to describe the photospheric flux contributions of V921\,Sco~B to the total flux
as function of wavelength:
$\frac{F_{\rm B}}{F_{\rm tot}}(\lambda) = \frac{F_{\rm B}}{F_{\rm tot}}(2~\mu\mathrm{m}) + s \cdot (\lambda - 2~\mu\mathrm{m})$.

We employ a Levenberg-Marquardt least square fitting procedure and
find the best-fit solution by minimizing the likelihood-estimator 
$\chi_{r}^2 = \chi_{r,V}^2 + \chi_{r,\Phi}^2$, where 
$\chi_{r,V}^2$ and $\chi_{r,\Phi}^2$ are the reduced least square between the measured and model
visibilities and closure phases, respectively.
The parameter uncertainties have been estimated 
using the bootstrapping technique.

Given that the absolute calibration of the AMBER+FINITO HR observations is not 
very reliable due to the use of long integration times and a potential residual 
phase jitter, we decided to recalibrate the absolute visibility level
using the detailed continuum model discussed in Sect.~\ref{sec:continuum} (Model DISK).
Since the HR-data was recorded 2010, we extrapolated the orbital
motion of the companion to this later epoch ($\rho=25.41$~mas, $\Theta=339^{\circ}$),
but note that the continuum geometry has only a marginal effect on our
model-fitting results in the strong {\BrG}-line of V921\,Sco.

\begin{figure*}
  \centering
  \includegraphics[angle=0,scale=1]{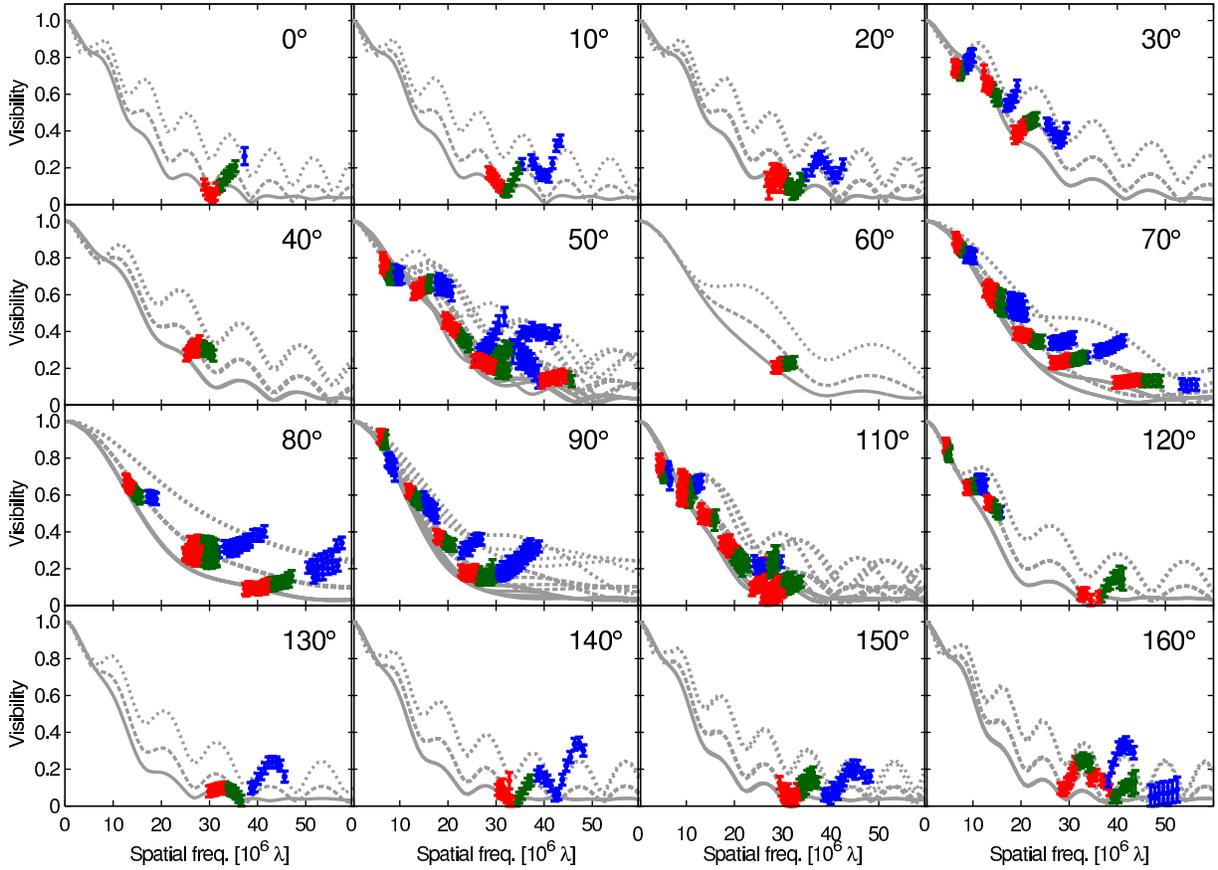} \\
  \caption{
    VLTI/AMBER visibilities (LR-HK mode) of V921\,Sco in the $H$- and $K$-band,
    plotted as function of spatial frequency.  The different panels represent position angle bins,
    covering 10$^{\circ}$ each.
    The grey model curves correspond to our best-fit temperature gradient disk model (DISK)
    and were computed for the lower (1.45~$\mu$m, dotted line), central (2~$\mu$m, dashed line), and 
    upper (2.55~$\mu$m, solid line) part of the wavelength range.
    For clarity, the data points are color-coded according to the wavelength of the corresponding spectral channel
      (blue: $1.4 \leq \lambda < 1.9~\mu$m; 
      green: $1.9 \leq \lambda < 2.15~\mu$m; 
      red: $2.15 \leq \lambda < 2.5~\mu$m).
  }
  \label{fig:modelVIS}
\end{figure*}

The DISK model provides a significant improvement compared to the
GAUSS model, which also reflects in an improved $\chi^2_r$-values of
1.86 (versus 4.88) for the 2008 data and 3.02 (versus 3.39) 
for the 2009 data,respectively.
In contrast to the GAUSS model, the DISK model also provides a physically
motivated parameterization for the wavelength-dependent changes in the 
source geometry, allowing us to fit all wavelength channels simultaneously.
The resulting best-fit parameters for the various models are listed in Tab.~\ref{tab:modelfitting}.
We plot in Fig.~\ref{fig:modelVIS} the 
measured observables versus the prediction of the DISK model.
Besides the spatially resolved emission around the primary component, we also checked whether
the fit results can be improved by including spatially extended emission around the northern 
(secondary) component, but were not able to find a significant improvement.

\begin{figure*}
  \centering
  \includegraphics[angle=0,scale=0.8]{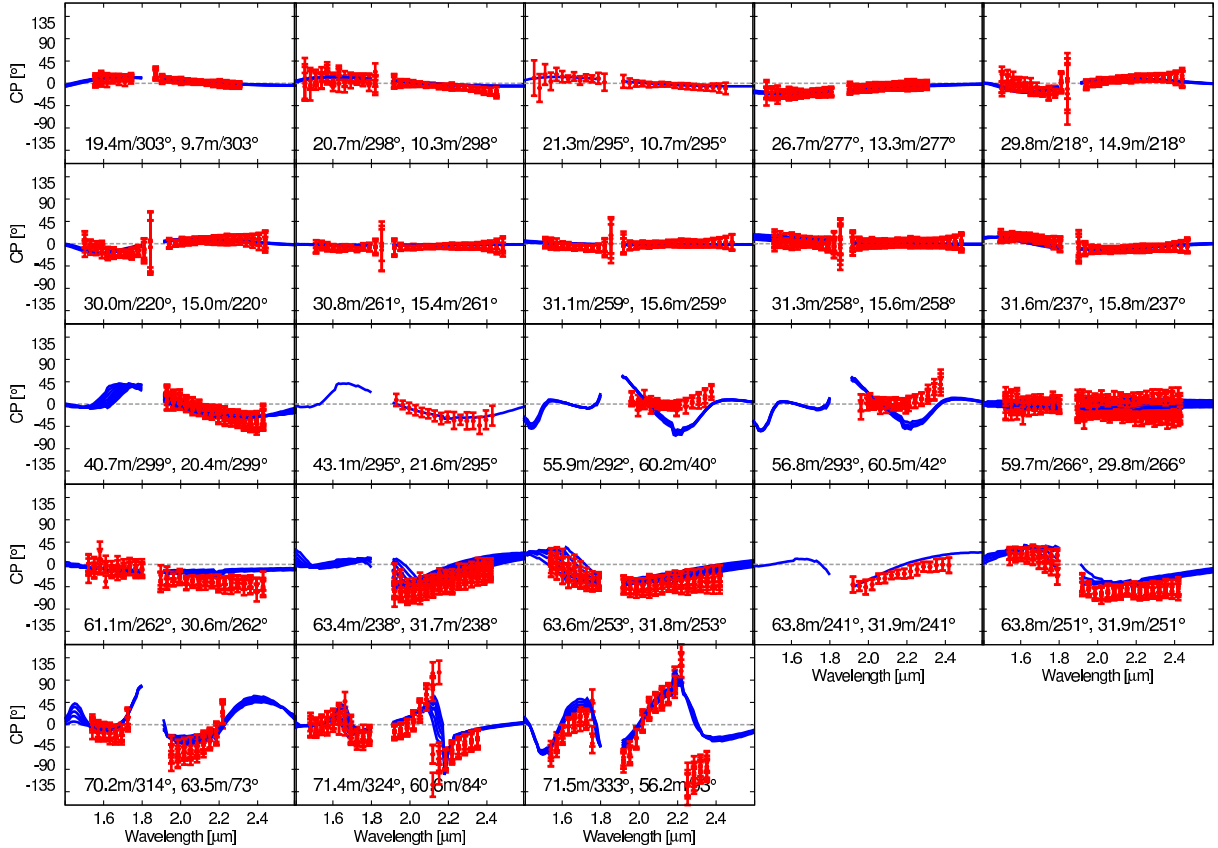} \\
  \caption{
    VLTI/AMBER closure phases and model closure phases for our best-fit model DISK.
    In each panel, we give the projected length and PA for two of the three employed baselines
      (the values for the third baseline are given by the closure relation).
  }
  \label{fig:modelCP}
\end{figure*}

\begin{deluxetable*}{lcccccccccccccccc}
\tabletypesize{\scriptsize}
\tablecolumns{16}
\tablewidth{0pc}
\tablecaption{Model-fitting results for the VLTI/AMBER continuum observations (Sect.~\ref{sec:continuum})\label{tab:modelfitting}}
\tablehead{
                         &                 & \multicolumn{5}{c}{Secondary star}                                        && \multicolumn{6}{c}{Circumprimary disk}                                          &                     &                       & \\
\cline{3-7} \cline{9-14}
\colhead{Epoch} & \colhead{Model}  & \colhead{$\Theta$}        & \colhead{$\rho$}          & \colhead{$\theta_{\rm B}$}  & \colhead{$\frac{F_{\rm B}}{F_{\rm tot}}$}
                                                                                                                & \colhead{$s$}            && \colhead{$\theta_{\rm A}$} & \colhead{$i$}       &  \colhead{$\varphi$}       &  \colhead{$R_{\rm in}$} & \colhead{$R_{\rm out}$} & \colhead{$q$}   &   \colhead{$\chi^{2}_{\rm r,V}$} & \colhead{$\chi^{2}_{\rm r,\Phi}$} & \colhead{$\chi^{2}_{\rm r}$} \\
                &                  & \colhead{[\deg]}          & \colhead{[mas]}           & \colhead{[mas]}            & \colhead{(at $2~\mu\mathrm{m}$)} & \colhead{[$\mu$m$^{-1}$]} && \colhead{[mas]}           & \colhead{[\deg]}    & \colhead{[\deg]}       & \colhead{[mas]}   & \colhead{[mas]}      &     &      &    &  \\
}
\startdata
2008      & GAUSS                         & $353.8$       & $25.0$          & $\leq 0.2$       & $0.054$          & $-0.0056$        && $7.5$         & $50.3$        & $147.8$      & --                & --          & --           & 4.30    & 3.09                   & 4.88\\
          &                               & $\pm1.6$      & $\pm0.8$       &                   & $\pm0.018$       & $\pm0.019$       && $\pm0.2$      & $\pm1.9$      & $\pm4.3$     &                     &             &              &       &                        & \\
2008      & DISK                          & $354.0$       & $24.9$          & $\leq 0.7$      & $0.066$        & $-0.051$   && --             & $48.8$  & $145.0$    & $1.59$ & $>9.7$      & $0.36$  & 1.86                & 1.86                   & 1.86\\ 

          &                               & $\pm 1.0$     & $\pm 0.5$       &                 & $\pm 0.04$     & $\pm 0.01$ &&                  & $\pm 4$  & $\pm 8.4$    & $\pm0.25$ &      & $\pm0.08$  &                   &                    & \\
  \hline
2009      & GAUSS                         & $347.3$       & $25.5$         & $\leq 0.2$\tablenotemark{a}     & 0.054\tablenotemark{a}         & -0.0056\tablenotemark{a}      && 7.5\tablenotemark{a}       & 50.3\tablenotemark{a}        & 147.8\tablenotemark{a}     & --          & --           & --    & 4.17            & 1.78                   & 3.39\\
          &                               & $\pm1.0$      & $\pm1.2$       &                  &                   &                 &&                &                &               &                 &                        & \\ 
2009      & DISK                          & $347.4$     & $25.4$         & 0.7$^a$        & $0.086$      & -0.051$^a$     && --             & 48.8$^a$   & 145.0$^a$    & 1.59$^a$    & 9.7$^a$     & 0.36$^a$ & 3.26             & 2.52                   & 3.02 \\
          &                               & $\pm 1.1$   & $\pm0.7$       &                & $\pm0.012$   &                &&                &   &     &    &     &        &                 &                        &  
\enddata
\tablenotetext{a}{In our fitting procedure, this parameter was kept fixed.}
\end{deluxetable*}

\section{Results: {\BrG}-line geometry}
\label{sec:line}

\subsection{2-D photocenter analysis}
\label{sec:photocenter}

\begin{figure*}
  \centering
  $\begin{array}{c@{\hspace{5mm}}c}
    \includegraphics[angle=0,scale=0.7]{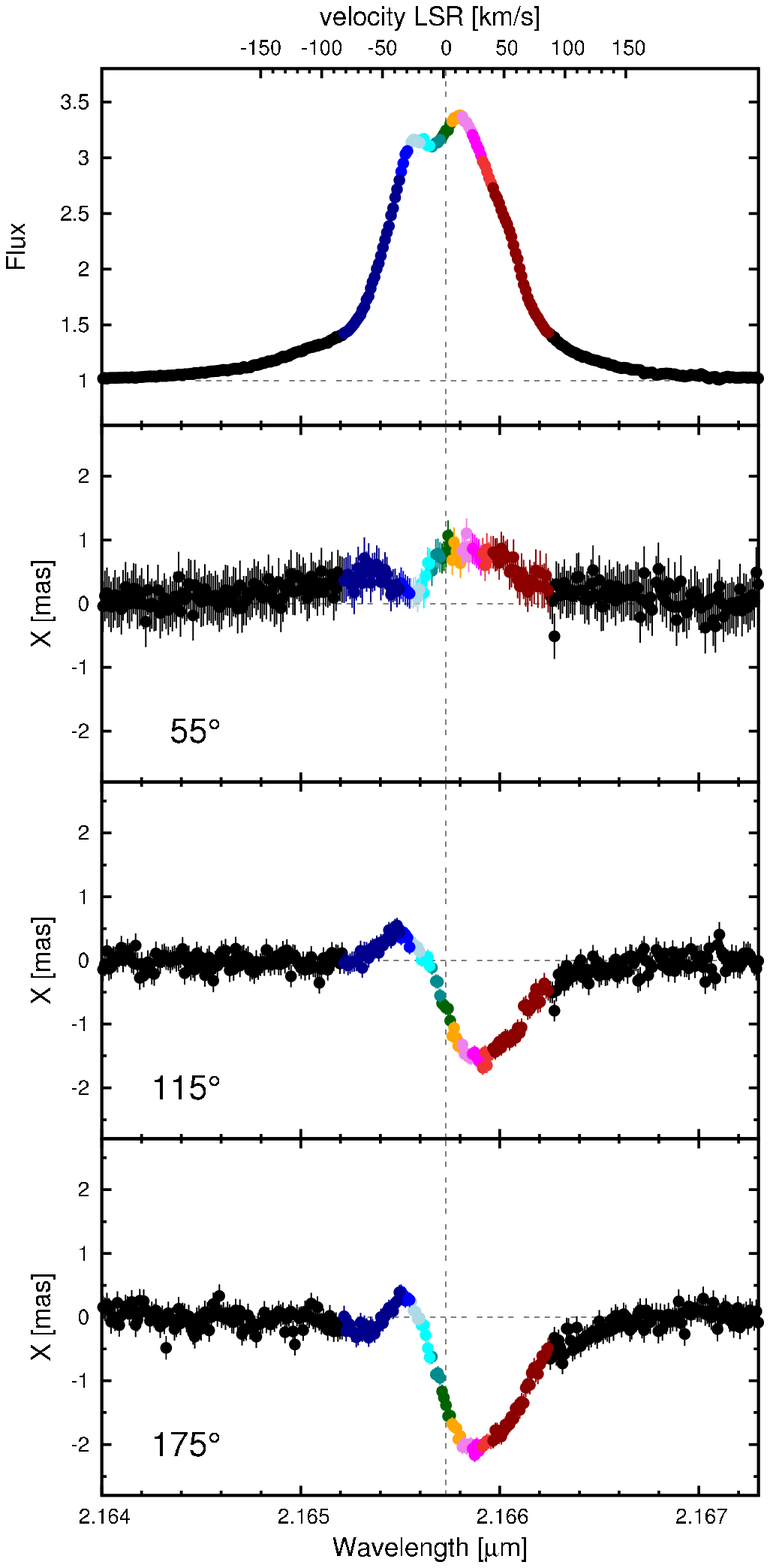} &
    \begin{minipage}{10cm}
      \vspace{-13.7cm}
      \includegraphics[angle=0,scale=0.9]{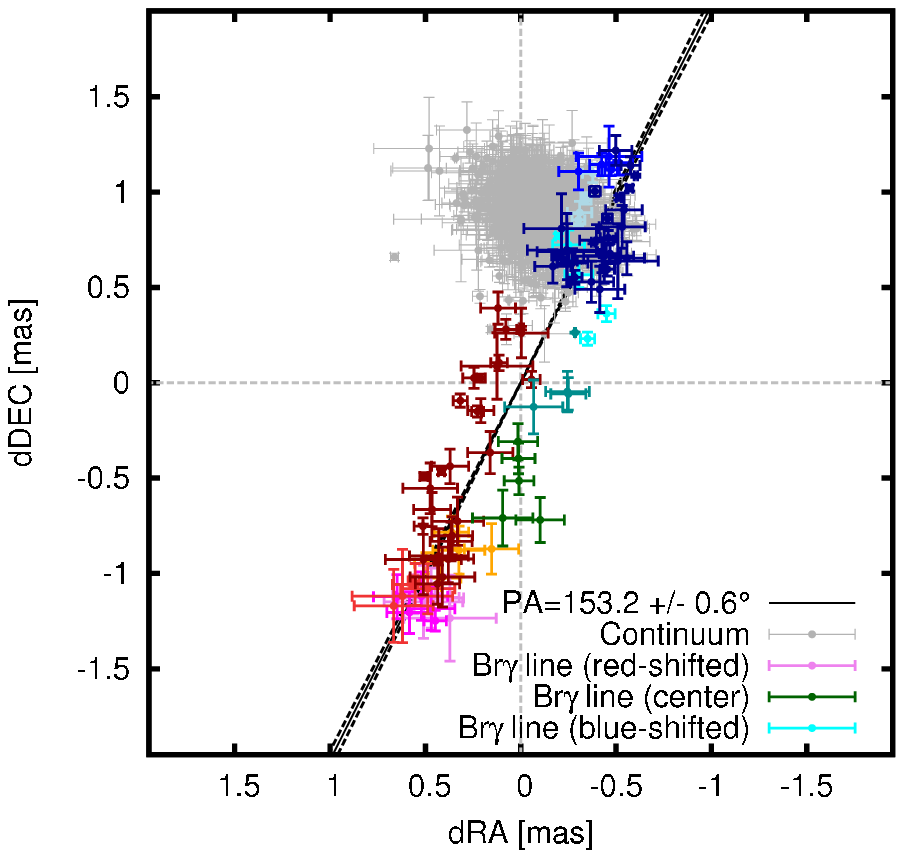}
    \end{minipage}
  \end{array}$
  \caption{
    {\it Left:} Spectrum (top) and spectro-astrometric signal (2nd-4th panel from top) derived from our VLT/CRIRES ($R=100,000$) observations.
    {\it Right:} Derived 2-D photocenter displacements, where the different spectral channels 
    are represented by different colors, matching the color coding in the left panel.
    From the photocenter displacements measured in the different spectral channels,
    we derive the disk rotation plane to $153.2 \pm 0.6^{\circ}$.
    The continuum spectral channels (grey points) are offset with respect to the center of light of the line emission
    (origin of the coordinate system) due to the presence of the companion.
  }
  \label{fig:photocenterCRIRES}
\end{figure*}

For a first, qualitative interpretation (Sect.~\ref{sec:photocenter})
of the gas kinematics, we derive the photocenter displacement of the
line-emitting region with respect to the continuum emission using our CRIRES spectro-astrometric data.
For this purpose, it is necessary to separate the line spectro-astrometric signal 
from the underlying continuum contributions.
To determine the continuum level, we fit a high-order polynomial function $X_c$ 
to the astrometric signal derived from the continuum channels.
This function is then subtracted from the astrometric signal in the line channels
and weighted by the continuum-to-line flux ratio $F_{c}/F_{l}$:
\begin{equation}
  X_{l}(v) = \left( X(v) - X_c(v) \right) \left( 1+F_{c}(v)/F_{l}(v) \right)
\end{equation}

The derived photocenter of the line velocity channels is significantly offset with 
respect to each other (clearly indicating the gas kinematics),
and also shows a displacement with respect to the 
photocenter of the continuum channels (Fig.~\ref{fig:photocenterCRIRES}).  
This displacement reflects the fact that the center of gravity 
of the continuum emission does not coincide with the location of the primary
star, but is displaced towards the position of the companion.
We calculate the center of gravity both for the line channels 
and the continuum channels and compute their relative displacement vector $\vec{v}_{c-l}$,
which is related to the companion separation $\rho$, position angle $\Theta$, 
and the flux ratio $\frac{F_{\rm B}}{F_{\rm tot}}$ by
\begin{equation}
  \vec{v}_{c-l} = \rho \frac{F_{\rm B}}{F_{\rm tot}} \left( \begin{array}{c} \sin \Theta \\ \cos \Theta \end{array} \right).
\end{equation}
Based on our continuum model fits from 2008 and 2009 (Tab.~\ref{tab:modelfitting}), 
we fix $\rho=25.13$~mas and then determine 
the companion position angle $\Theta$
and the flux ratio $F_B/F_{\rm tot}$
from the CRIRES observations.
Applying this procedure yields $\Theta = 353 \pm 9^{\circ}$
(which is in reasonable agreement with the astrometry determined 
with AMBER, Tab.~\ref{tab:modelfitting})
and $F_B/F_{\rm tot} = 0.053$, which is lower than the 
flux ratio determined with interferometry for epochs 2008 and 2009, 
possibly indicating variability.

The photocenters in the blue- and red-shifted line wings are displaced
in opposite direction with respect to each other.  Also, the highest gas velocities
(dark red points and dark blue points in Fig.~\ref{fig:photocenterCRIRES})
emerge from smaller stellocentric distances than intermediate velocities,
which suggests a rotation-dominated velocity profile.
From the distribution of the individual photocenter offsets, we derive the disk plane orientation
to $153.2\pm0.6^{\circ}$.

In contrast to CRIRES, AMBER is able to spatially resolve the geometry of the
line-emitting region, 
entering a regime where higher-order geometric effects are probed.
Therefore, in the following two sections, we will employ a quantitative
modeling in order to interpret the combined spectro-astrometric
and spectro-interferometric observations.

\subsection{Model fitting: Keplerian disk}
\label{sec:linemodeldisk}

\begin{figure*}
  \centering
  $\begin{array}{c}
    \includegraphics[angle=0,scale=0.55]{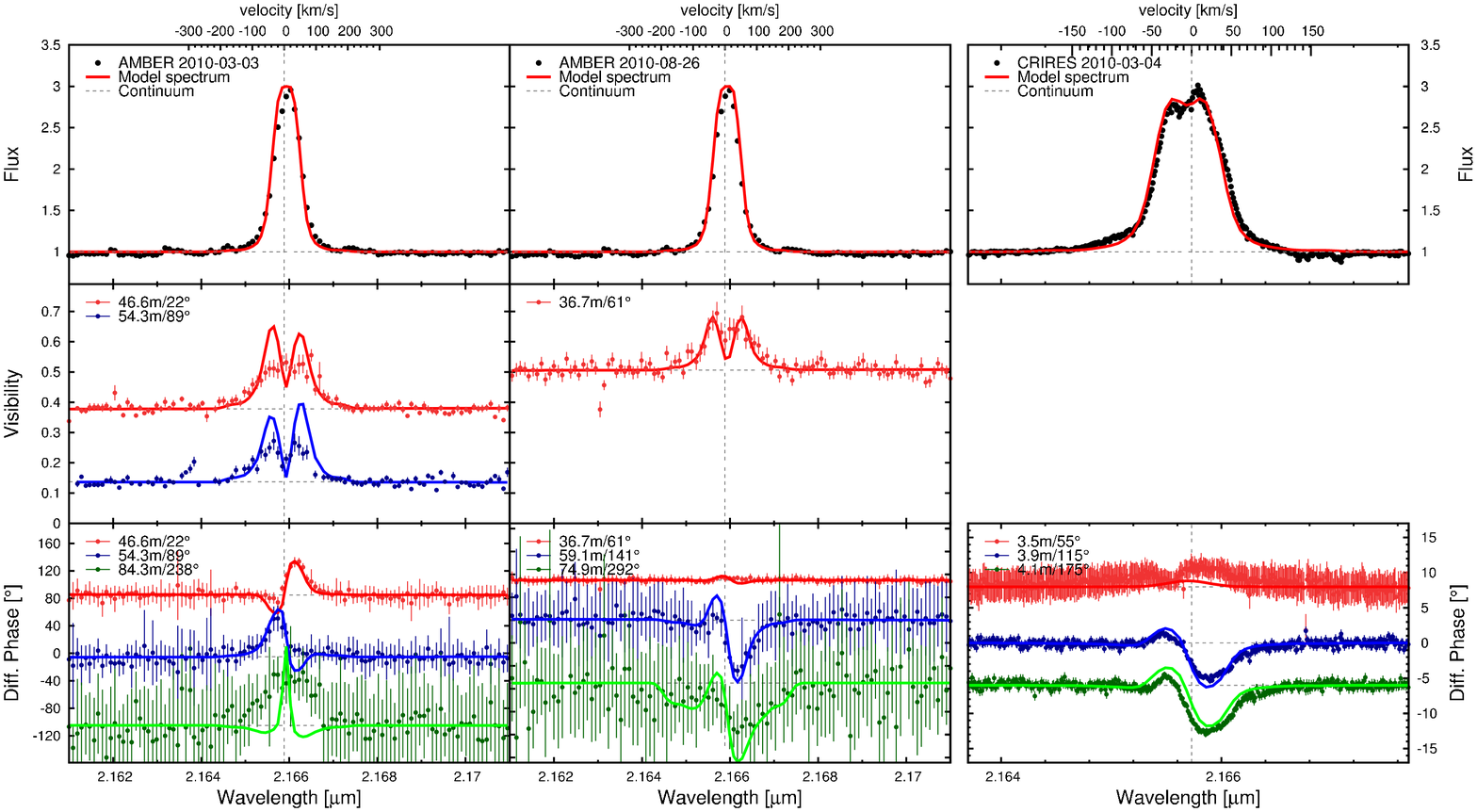}\\[3mm]
    \includegraphics[angle=0,scale=0.412]{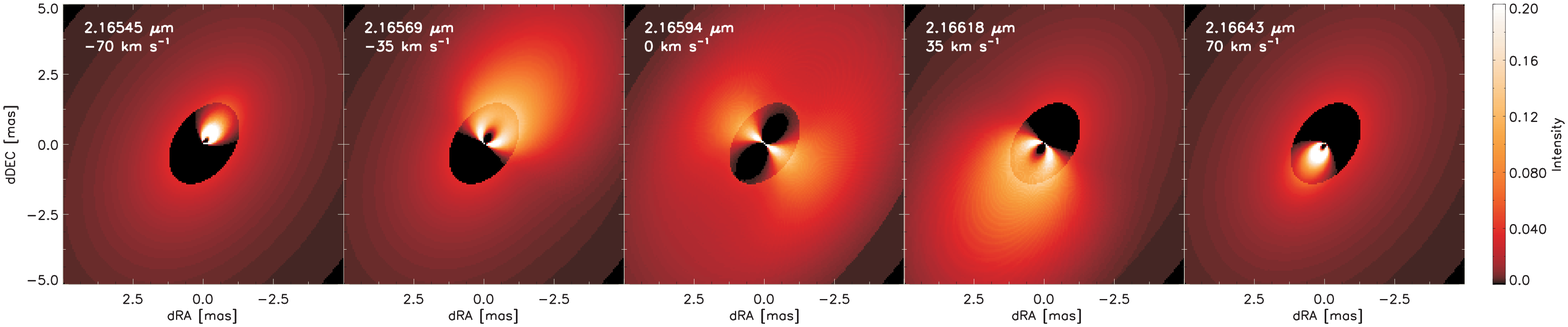}
  \end{array}$
  \caption{
    {\it Top:} Spectra {\it (1st row)}, visibilities {\it (2nd row)}, and differential phases {\it (3rd row)} derived
    from our AMBER and CRIRES {\BrG} observations, overplotted with the Keplerian disk model (Sect.~\ref{sec:linemodeldisk}).
    For each interferometric baseline, we give the corresponding projected baseline length and PA in the label.
    In order to make the visibility profiles in the left panel clearly distinguishable, 
    we added an offset of 0.2 to the visibility corresponding to the baseline 46.6m/22$^{\circ}$.
    {\it Bottom:} To illustrate our kinematical model we show channel maps for some representative wavelengths, 
    including the {\BrG}-line and the continuum emission.  
    The continuum-emitting disk is truncated at a distance of 1.59~mas, while the
    {\BrG}-emitting material extends inwards to a few stellar radius ($\lesssim 0.5$~mas).
    The companion star V921\,Sco~B is also included in these simulations, 
    but located outside of the FOV in these model images.
  }
  \label{fig:modelKEPLER}
\end{figure*}

\begin{deluxetable}{lccc}
\tabletypesize{\scriptsize}
\tablecolumns{4}
\tablewidth{0pc}
\tablecaption{Model-fitting results for the VLTI/AMBER (HR-K) and VLT/CRIRES {\BrG}-line data (Sect.~\ref{sec:line}).\label{tab:linemodelfitting}}
\tablehead{ 
\colhead{Parameter} &  &  & \colhead{Value} 
}
\startdata
  Inner emission radius            & $R_{\rm in}$       & [mas]       & $\lesssim 0.5$ \\  
  Outer emission radius            & $R_{\rm out}$      & [mas]       & $5.8 \pm 0.3$ \\  
  Position angle                   & $\theta$         & [$\deg$]    & $145.0 \pm 1.3$ \\
  Inclination                      & $i$              & [$\deg$]    & $54 \pm 2$ \\
  Stellar mass                     & $M_{\star}$        & [$M_{\sun}$] & $(5.4 \pm 0.4) \cdot (d/1150~\mathrm{pc})$ \\
  Radial intensity index           & $q$              &             & $-1.0 \pm 0.1$ \\
  Opacity index                    & $\kappa$         &             & $0.5 \pm 0.2$
\enddata
\end{deluxetable}

Given the indications for a rotation-dominated velocity field provided
by our model-independent photocenter analysis (Sect.~\ref{sec:photocenter}), 
we first test whether a Keplerian velocity field might reproduce
our spectroscopic, spectro-astrometric and spectro-interferometric data quantitatively.

For this purpose, we employ a Keplerian disk model ($v(r) = \sqrt{GM_{\star}/r}$), 
which we have already successfully applied to AMBER high spectral dispersion 
data on the classical Be star $\beta$\,CMi \citep{kra12a}.
The model assumes that the line-emitting gas is optically thin and 
located in a geometrically thin disk,
where the radial intensity profile is parameterized with a power-law ($I_{l}(r) \propto r^{q}$).
The gas emission extends from an inner truncation radius ($R_{\rm in}$)
to an outer truncation radius ($R_{\rm out}$), which we parameterize
with a Fermi-type function in order to avoid artifical edges
(see Eq.~1 in \citealt{kra08a}, where the width of the truncation region $\epsilon$
was chosen to 0.1). 
In our study on $\beta$\,CMi \citep{kra12a}, we investigated the 
DP signatures induced by photospheric absorption and found 
that even for a relatively extended stellar surface (equatorial radius 
of 0.36 mas), the induced DPs are $<0.5^{\circ}$.
For V921\,Sco, the influence of photospheric absorption is likely
$\sim 50$-times smaller due to the smaller apparent stellar radius ($\sim 0.07$~mas)
and the significantly larger equivalent width of the {\BrG} emission line.
Therefore, these signatures are about four orders of magnitude
smaller than the disk kinematical signatures and can be safely
neglected in our modeling process.

As discussed in Sect.~\ref{sec:photocenter}, the elliptical distribution of the derived 
photocenter vectors might indicate opacity effects, which cause the more 
distant parts of the disk to appear fainter than the disk parts facing the observer. 
In order to include this effect in our model, we assume that the disk is embedded in an medium
of constant density.  After a distance $x$ traveled, the emitted intensity $I_0$ is then reduced
to $I(x)=I_0 \exp(-\kappa x)$.
Combining this line-emission model with our best-fit continuum disk+companion model 
(DISK, Tab.~\ref{tab:modelfitting}) allows us to produce model channel maps (Fig.~\ref{fig:modelKEPLER}, {\it bottom}),
from which we compute line profiles, visibilities, and DPs for comparison with our data (Fig.~\ref{fig:modelKEPLER}, {\it top}).

As line broadening mechanisms, we include thermal and turbulent
Doppler broadening \citep{pie07}
\begin{equation}
  \Delta V= \sqrt{\frac{2 kT_{\rm gas}}{m} + v_{\rm turb}^2},
\end{equation}
where $v_{\rm turb}$ is the turbulent velocity,
which is typically negligible for Herbig~Ae/Be stars
($\lesssim 0.5$~km\,s$^{-1}$, e.g.\ \citealt{pie07}).
In the disk surface layer, the gas temperature $T_{\rm gas}$ will 
be significantly higher than the dust temperature at a given
stellocentric radius. We assume $T_{\rm gas} = 5~T_{\rm dust}$,
as suggested by the thermally decoupled radiative-hydrodynamics
simulations by \citet{thi11}, but note that the precise value 
affects mainly the line width and does not affect our 
general conclusions (see also Sect.~\ref{sec:interpFIRE}).

Free parameters in our model are the mass of the central star $M_A$, 
the inner and outer disk radius ($R_{\rm in}$, $R_{\rm out}$), 
the radial intensity power-law index $q$, the disk inclination $i$, 
and the opacity index $\kappa$.

We vary these parameters systematically on a parameter grid
and select the best-fit model with the best 
$\chi^2_r = \chi^2_{r,V} + \chi^2_{r, \phi} + \chi^2_{r,F}$, where
$\chi^2_{r,\phi}$ and $\chi^2_{r,F}$ are the reduced chi-squared between 
the model and measured different phase and spectrum, respectively.

Remarkably, this simple disk model can reproduce important features in our data, in particular:

\begin{itemize}
\item[(a)] Our model can reproduce the double-peaked {\BrG}-line profile 
  (Fig.~\ref{fig:modelKEPLER}, {\it 1st panel}) reasonably well.
  One line profile characteristic, which is not reproduced by our model
  concerns the weak asymmetry observed in {\BrG}-line profile,
  with a slightly stronger red-shifted line wing.

\item[(b)] Our AMBER observations (Fig.~\ref{fig:modelKEPLER}, {\it 2nd panel}) 
  reveal M-shaped visibility profiles on two of three baselines (36.7m/$61^{\circ}$ and 53.4m/$89^{\circ}$),
  indicating that the angular extension of the line-emitting region increases for low gas velocities.
  This effect is expected for a rotation-dominated velocity field, 
  where the azimuthal velocity decreases as function of radius
  and the visibility profile shape is reasonably well reproduced by our model for these baselines.
  On the third baseline (46.6m/$22^{\circ}$), no central visibility drop has been observed.
  We suspect that this effect is related to more subtle radiative transfer or line broadening effects,
  which are not included in our simplistic model.
  The visibilities at low gas velocities (i.e.\ in the line center) 
  would be most sensitive to such small-order effects and we leave it to 
  future studies using full radiative transfer modeling to investigate this in more detail.

\item[(c)] The measured AMBER and CRIRES DPs (Fig.~\ref{fig:modelKEPLER}, {\it 3rd panel})
  show very interesting signatures, including asymmetric S-shaped signatures
  and V-shaped signatures, which appear sometimes in the blue- and
  sometimes in the red-shifted line wing.
  These signatures constrain the disk kinematics primarily, but contain
  also contributions from the continuum photocenter 
  displacement caused by the companion star.
  Our combined binary star plus Keplerian disk rotation model reproduces the
  large variety of signatures reasonably well.
  The strongest residuals between the model and the data are observed at the longest 
  AMBER baselines, where the data is extremely sensitive to small-order kinematical effects,
  but where the SNR is also reduced due to the lower fringe contrast.
\end{itemize}

The parameters corresponding to the best-fit model are listed in 
Tab.~\ref{tab:linemodelfitting} and will be discussed in more detail 
in Sect.~\ref{sec:interpdisk}.

We also investigated whether a significant fraction of the
{\BrG} emission might be associated with the secondary star (V921\,Sco~B)
instead of the circumprimary disk, possibly indicating active accretion
from the circumbinary disk onto the secondary star.
Considering the strong measured displacement of the line photocenter
in the direction towards the primary (Fig.~\ref{fig:photocenterCRIRES}, {\it right})
and the good quantitative agreement of the displacement vector with the 
flux-weighted binary astrometry vector (Sect.~\ref{sec:photocenter}), 
it is clear that the secondary is not the dominant
{\BrG}-emitting component.
In order to further quantify these constraints, we introduced 
the fraction of {\BrG} emission associated with the secondary star to the total {\BrG}-flux 
$(F_{\rm B}/F_{\rm tot})_{\mathrm{Br}\gamma}$ as an additional parameter in our kinematical model and find that 
$(F_{\rm B}/F_{\rm tot})_{\mathrm{Br}\gamma} < 0.11$.

\section{Interpretation}
\label{sec:interp}

\subsection{Spectroscopy}
\label{sec:interpFIRE}

\begin{figure*}
  \centering
 \includegraphics[angle=0,scale=0.53]{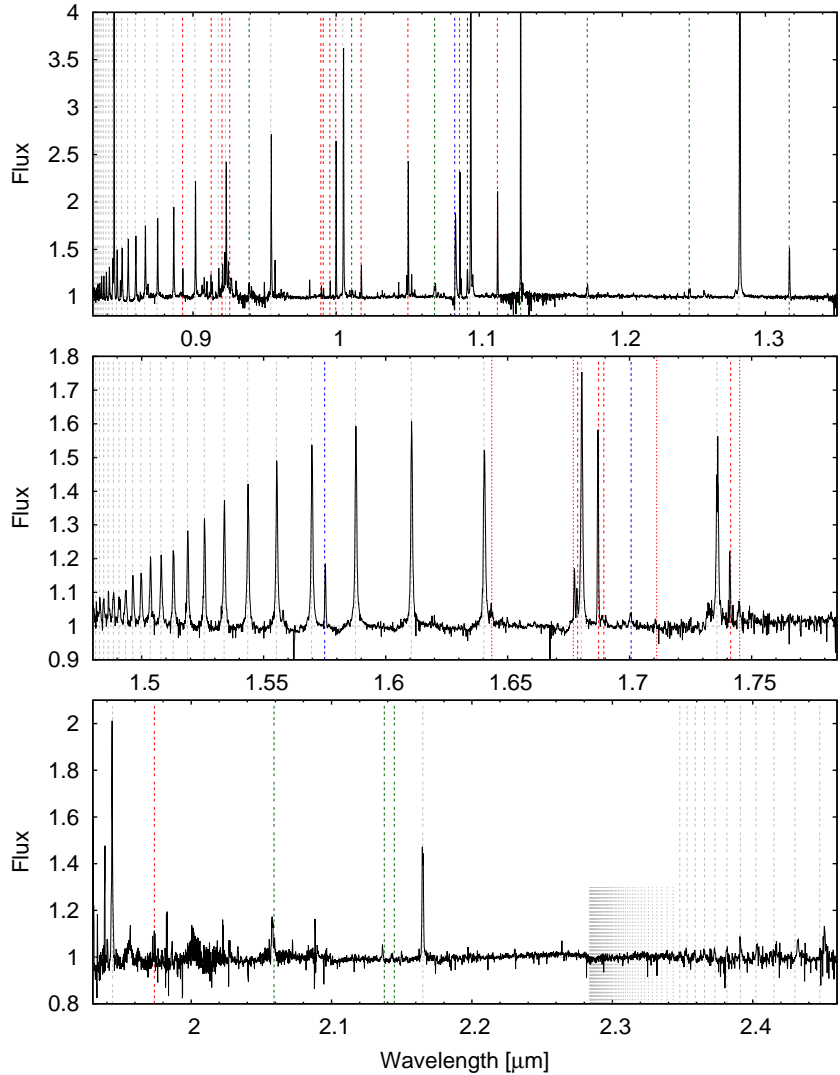} \\
  \caption{
    Near-infrared $J$-band ({\it top}), $H$-band ({\it middle}), and $K$-band ({\it bottom}) spectrum 
    recorded with the Magellan/FIRE spectrograph.
    The spectrum is dominated by strong hydrogen emission lines from the Paschen, Brackett, and Pfund series
    ({\it grey lines}) and metallic lines from \ion{Fe}{2} ({\it red, dashed lines}),
    [\ion{Fe}{2}] ({\it red, dotted lines}), \ion{Mg}{2}, \ion{N}{1}, \ion{Al}{2}, \ion{C}{1}, \ion{O}{1} ({\it green lines}).
    A complete list of the identified lines can be found in Tab.~\ref{tab:FIRElines}.
    For the Pfund lines, we show besides the clearly identified transitions (Pf14-24, {\it grey dashed lines}) also 
    the higher-order line transitions (Pf25-100, {\it grey dotted lines}) in order to mark the location of the Pfund discontinuity.
  }
  \label{fig:FIRE}
\end{figure*}

Our FIRE observations (Fig.~\ref{fig:FIRE}) reveal a very rich near-infrared spectrum.  
Similar to the optical spectrum \citep{bor07}, the near-infrared regime is dominated 
by strong hydrogen line emission.
As listed in Tab.~\ref{tab:FIRElines}, we detect 27 lines from the Brackett series ({\BrG} to Br29),
24 lines from the Paschen series (Pa$\alpha$ to Pa24), and at least 10 lines from the Pfund series (Pf14-24).
Given that the recombination physics of hydrogen is well known, 
one can use these line decrements in order to derive information
about physical conditions of the line-emitting gas.

In order to model the line flux in the optically thin approximation, 
we extracted from \citet{sto95} the line emissivities $\epsilon_{ul}$ 
for Case~B recombination 
for all Balmer ($l=3$), Brackett ($l=4$), and Paschen ($l=5$) hydrogen 
line transitions $(u-l)$ with $u\leq25$.
The line emissivities include collisional transitions,
which makes our computation applicable even for relatively high gas densities.
Assuming that the emitting gas is isothermal ($T=10,000$~K)
and uniformly distributed in the volume $V$,
the received line flux at distance $d$ is then given by
\begin{equation}
  F_{u-l} = \frac{N_e N_p \epsilon_{ul} V}{4 \pi d^2},
\end{equation}
where $N_e$ and $N_p$ are the electron and proton density per unit volume.
In order to approximate the emitting volume $V$, we estimate from our
{\BrG} spectro-interferometric data the characteristic stellocentric emission radius.
For this purpose, we generate a velocity-integrated line map and measure
the half-light emission radius to $r_{\mathrm{line}}=3.0$~mas=3.5~AU (for $d=1.15$~kpc).
Assuming that the line-emitting gas is located in a hot disk surface layer
with a characteristic scale height $h$ of $h/r_{\mathrm{line}}=0.2$,
as suggested by the thermally decoupled simulations from \citet{thi11},
the emitting volume is given by $V=\pi r_{\mathrm{Br}\gamma}^{3} \cdot (h/r_{\mathrm{line}})$.
We employ an iterative procedure, where we start by assuming a low gas density,
which is then adjusted in order to reproduce the measured line fluxes.

\begin{figure*}
  \centering
  \includegraphics[angle=0,scale=1.2]{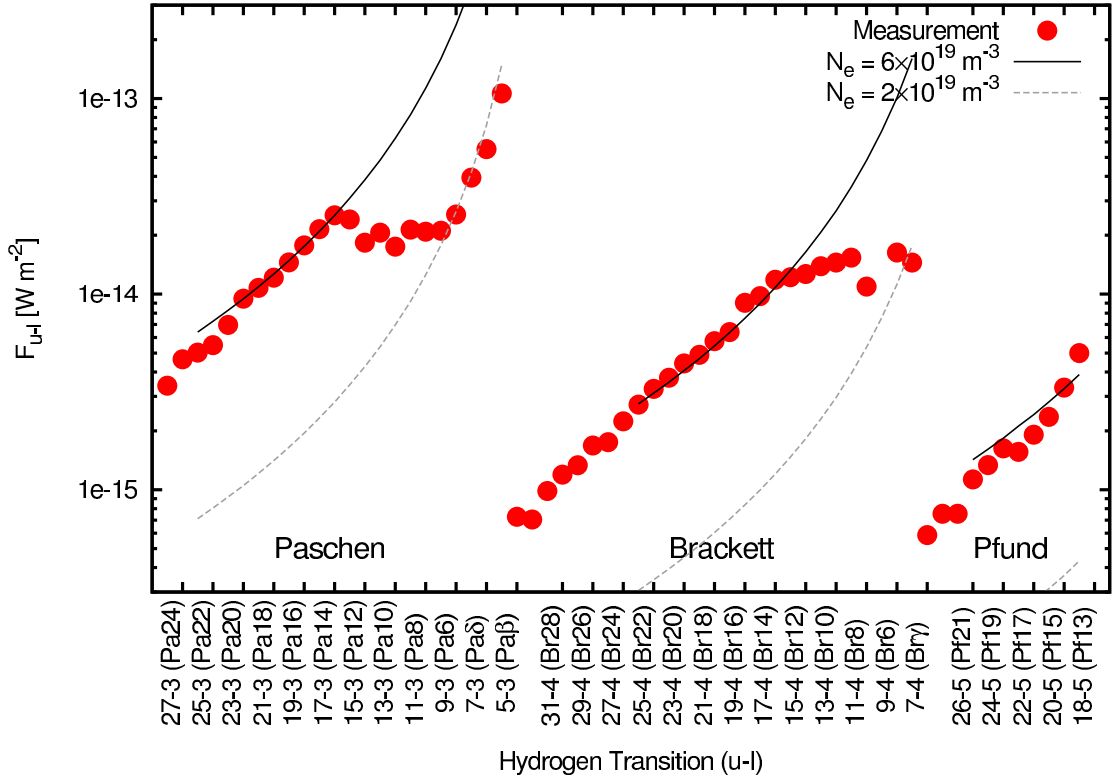} \\
  \caption{
    Line flux for the Paschen, Brackett, and Pfund decrements, 
    as derived from our V921\,Sco FIRE spectrum.
    The abscissa gives the upper ($u$) and lower ($l$) level for
    the corresponding hydrogen transition, where we label only
    about every other transition for space reasons.
    The measured line fluxes (points) have been corrected for reddening
    (Sect.~\ref{sec:obsFIRE}) and are overplotted with
    model recombination line fluxes for isothermal gas in 
    a disk surface layer with temperature of $T=10,000$~K
    and densities in the range $N_e = 2...6\times 10^{19}$\,m$^{-3}$
    (Sect.~\ref{sec:interpFIRE}).
  }
  \label{fig:lineflux}
\end{figure*}

Then we recompute the line emissivities and repeated the computation until convergence is reached.
As shown in Fig.~\ref{fig:lineflux}, we require rather high gas densities between
$N_e = N_p = 2\times 10^{19}$\,m$^{-3}$ and $6\times 10^{19}$m\,$^{-3}$,
which supports the scenario that the gas is located in a 
dense disk instead of a low-density halo.
For the high-level transitions ($u>16$, i.e.\ Pa14-24, Br13-29, and Pf12-23), 
the line fluxes can be reproduced well with a single gas density 
($N_e = 6\times 10^{19}$\,m$^{-3}$),
suggesting that the optical thin approximation is well justified.
The low-level transitions ($u<=16$, i.e.\ {\PaB}-Pa13, {\BrG}-Br12)
and more consistent with densities around $N_e = 2\times 10^{19}$\,m$^{-3}$,
which might indicating that these transitions originate from more 
extended disk regions, where the disk surface density is lower.
We would like to note that the derived densities depend to some extend also on
the assumed gas temperature.
Varying the temperature from the assumed $T=10,000$~K
to other realistic values (e.g.\ 5,000-20,000~K), would change the derived 
densities by a factor of $\sim 2-3$.
Based on our simplistic modeling, the derived density should be treated as an 
order-of-magnitude estimate and future radiation-hydrodynamic modeling will be required 
in order to fully exploit the rich information provided by our combined 
spectro-interferometric and spectroscopic dataset.

It is interesting to compare the derived electron density 
to the values predicted by state-of-the-art disk models.
For instance, for a typical disk around a Herbig~Ae star 
(e.g.\ \object{AB Aurigae}, 2.4~$M_{\sun}$) with a gas surface density of $10^{4}$~kg\,m$^{-2}$, 
the predicted midplane gas density is $\sim 6\times 10^{17}$\,m$^{-3}$ \citep{dul10}
at the dust sublimation radius (1.1~AU).
Our density measurement of $N_e = 2...6\times 10^{19}$\,m$^{-3}$
probes similar spatial scales in the V921\,Sco disk
and is one to two orders of magnitude higher than this value,
suggesting that the disk around V921\,Sco is exceptionally massive.
This result is in line with the study by \citet{hen98},
who determined the total gas mass in the millimeter cores
around 25~Herbig~Ae/Be and FU~Orionis stars and measured the highest 
total gas mass in the core around V921\,Sco ($M_{\mathrm{gas}}=40~M_{\sun}$; 
core size: {$27\times 27$\arcsec} or $\sim 30,000\times 30,000$~AU at 1.15~kpc).
Observations with our spectroscopic \& spectro-interferometric approach
on a larger sample of protoplanetary disks will be necessary in order 
determine whether the derived high gas density indicates an
intrinsic property of the V921\,Sco disk (possibly due to the
young age of the source) or simply reflects the predicted
disk surface density scaling law with stellar mass 
($\Sigma \propto \dot{M} \propto M_{\star}^{2}$, \citealt{cal04}).

Besides hydrogen recombination lines, other identified lines include 
\ion{Fe}{2}, [\ion{Fe}{2}], \ion{C}{1}, \ion{He}{1}, 
\ion{O}{1}, \ion{N}{1}, \ion{Mg}{2}, \ion{Al}{2}, which have also been identified
in the B[e] stars \object{CI~Cam} \citep{cla99} and \object{HD\,50138} \citep{jas92}, 
or in the LBV \object{$\eta$\,Car} \citep{dam98,smi01}.
Besides the total of 90 identified lines, we detect in our spectra a few dozen additional, yet unidentified lines
and we encourage detailed follow-up spectroscopic studies in order to investigate the rich chemistry of this source.

Our wavelength range also covers the CO first overtone bandheads between 2.3 and 2.4~$\mu$m,
which allows us to search for the CO first overtone bandheads, which we detect neither in 
absorption nor emission.
This is particularly interesting, since \citet{kra09c} suggested that these lines
provide a good diagnostic tool to distinguish between a pre- and post-main-sequence 
evolutionary phase in B[e] stars.  According to her computation, observable $^{13}$CO bandhead emission
can only be produced in evolved stars, but should be absent in pre-main-sequence B[e] stars.

\begin{figure}
  \centering
  \includegraphics[angle=0,scale=0.35,angle=270]{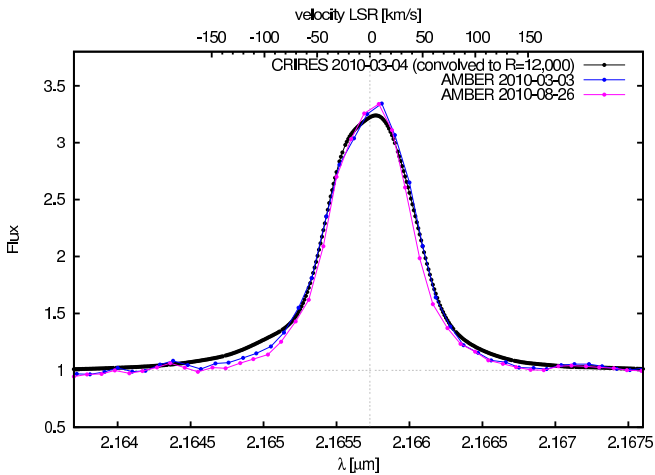} \\
  \caption{VLT/CRIRES and VLTI/AMBER {\BrG}-line spectra measured in 
      March and August 2010, which have been convolved to a
      common spectral resolution of $R=12,000$.}
    \label{fig:linevariability}
\end{figure}

\begin{deluxetable}{ccccc}
\tabletypesize{\scriptsize}
\tablecolumns{4}
\tablewidth{0pc}
\tablecaption{Lines identified in our V921\,Sco FIRE spectrum and measured line equivalent widths (EW) and integrated line fluxes $F$
(for lines in the $J$-band).\label{tab:FIRElines}}
\tablehead{
\colhead{Wavelength}    & \colhead{Line}   & \colhead{EW}   & \colhead{$F$}\\
\colhead{\[[$\mu$m]}    &                  & \colhead{[nm]} & \colhead{[$10^{-16}$~W\,m$^{-2}$]}\\
}
\startdata
0.8304        & Pa24          & -0.053\tablenotemark{a} & 14.7 \\
0.8312        & Pa23          & -0.073 & 20.1 \\
0.8321        & Pa22          & -0.079 & 21.6 \\
0.8332        & Pa21          & -0.087 & 23.7 \\
0.8343        & Pa20          & -0.110 & 29.7 \\
0.8357        & Pa19          & -0.153 & 41.0 \\
0.8372        & Pa18          & -0.175 & 46.4 \\
0.8390        & Pa17          & -0.199 & 52.2 \\
0.8411        & Pa16          & -0.242 & 62.7 \\
0.8436        & Pa15          & -0.300 & 76.6 \\
0.8447        & \ion{O}{1}    & -1.343 & 340.4 \\
0.8465        & Pa14          & -0.369 & 92.6 \\
0.8500        & Pa13          & -0.443 & 108.9 \\
0.8543        & Pa12          & -0.432 & 103.7 \\
0.8596        & Pa11          & -0.339 & 79.1 \\
0.8663        & Pa10          & -0.394 & 88.8 \\
0.8748        & Pa9           & -0.348 & 75.3 \\
0.8820        & \ion{O}{1}    & -0.012 & 2.5 \\   
0.8860        & Pa8           & -0.447 & 92.2 \\
0.8927        & \ion{Fe}{2}   & -0.097 & 19.5 \\   
0.9013        & Pa7           & -0.462 & 89.9 \\
0.9127        & \ion{Fe}{2}   & -0.074 & 13.9 \\   
0.9177        & \ion{Fe}{2}   & -0.169 & 31.2 \\   
0.9202        & \ion{Fe}{2}   & -0.121 & 22.1 \\   
0.9227        & Pa6           & -0.500\tablenotemark{a} & 90.8 \\
0.9256        & \ion{Fe}{2}   & -0.080\tablenotemark{a} & 14.4 \\   
0.9261        & \ion{O}{1}    & -0.070\tablenotemark{a} & 12.6 \\   
0.9392        & \ion{N}{1}    & -0.071\tablenotemark{a} & 12.3 \\    
0.9406        & \ion{C}{1}    & -0.060 & 10.4 \\
0.9543        & Pa$\epsilon$  & -0.660 & 110.1\\
0.9893        & \ion{Fe}{2}   & -0.045 & 6.9 \\   
0.9910        & \ion{Fe}{2}   & -0.029 & 4.5 \\   
0.9956        & \ion{Fe}{2}   & -0.070 & 10.7 \\   
0.9997        & \ion{Fe}{2}   & -0.624 & 94.4 \\   
1.0047        & Pa$\delta$    & -1.132 & 169.9 \\
1.0108        & \ion{Al}{2}   & -0.043\tablenotemark{a} & 6.4 \\   
1.0174        & \ion{Fe}{2}   & -0.143 & 21.0 \\   
1.0501        & \ion{Fe}{2}   & -0.549 & 76.6 \\   
1.0688        & \ion{C}{1}    & -0.163 & 22.2 \\    
1.0829        & \ion{He}{1}   & -0.760 & 101.7 \\   
1.0862        & \ion{Fe}{2}   & -0.503 & 67.1 \\   
1.0917        & \ion{He}{1}   & -0.207 & 27.4 \\   
1.0935        & Pa$\gamma$    & -1.793\tablenotemark{a} & 237.2 \\
1.1127        & \ion{Fe}{2}   & -0.468 & 60.6 \\   
1.1290        & \ion{O}{1}    & -1.874\tablenotemark{a} & 238.9 \\   
1.1755        & \ion{C}{1}    & -0.117 & 14.3 \\   
1.2467        & \ion{N}{1}    & -0.083\tablenotemark{a} & 9.6 \\   
1.2815        & Pa$\beta$     & -4.030\tablenotemark{a} & 457.5 \\
1.3166        & \ion{O}{1}    & -0.321 & 35.6    
\enddata
\tablenotetext{a}{These lines suffer from line-blending, which introduces 
  some uncertainty to the derived EW values.}
\tablenotetext{b}{Due to the high $K$-band continuum flux and line brightness, 
  these lines likely reached the non-linearity regime of the FIRE detector,
  resulting in an underestimation of the real EW.}
\end{deluxetable}

\begin{deluxetable}{ccccc}
\tabletypesize{\scriptsize}
\tablecolumns{4}
\tablewidth{0pc}
\tablecaption{{\it Continued:} Table~\ref{tab:FIRElines} for lines in the $H$- and $K$-band.}
\tablehead{
\colhead{Wavelength}    & \colhead{Line}   & \colhead{EW}   & \colhead{$F$}\\
\colhead{\[[$\mu$m]}    &                  & \colhead{[nm]} & \colhead{[$10^{-16}$~W\,m$^{-2}$]}\\
}
\startdata
1.4798        & Br29          & -0.031\tablenotemark{a} & 3.1 \\
1.4811        & Br28          & -0.030\tablenotemark{a} & 3.0 \\
1.4827        & Br27          & -0.042\tablenotemark{a} & 4.2 \\
1.4844        & Br26          & -0.051 & 5.1 \\
1.4863        & Br25          & -0.057 & 5.7 \\
1.4884        & Br24          & -0.072 & 7.3 \\
1.4907        & Br23          & -0.075\tablenotemark{a} & 7.5 \\
1.4934        & Br22          & -0.096 & 9.6 \\
1.4963        & Br21          & -0.117 & 11.7 \\
1.4997        & Br20          & -0.141 & 14.1 \\
1.5035        & Br19          & -0.161 & 16.1 \\
1.5078        & Br18          & -0.191 & 19.1 \\
1.5129        & Br17          & -0.212 & 21.1 \\
1.5188        & Br16          & -0.250 & 24.8 \\
1.5256        & Br15          & -0.279 & 27.6 \\
1.5338        & Br14          & -0.395 & 38.9 \\
1.5435        & Br13          & -0.430 & 42.1 \\
1.5552        & Br12          & -0.524\tablenotemark{a} & 51.0 \\
1.5696        & Br11          & -0.544 & 52.6 \\
1.5750        & \ion{Fe}{2}   & -0.105 & 10.1 \\   
1.5876        & Br10          & -0.570 & 54.7 \\
1.6105        & Br9           & -0.631 & 59.9 \\
1.6403        & Br8           & -0.667\tablenotemark{a} & 62.5 \\
1.6435        & [\ion{Fe}{2}] & -0.019\tablenotemark{a} & 1.8 \\   
1.6769        & [\ion{Fe}{2}] & -0.076\tablenotemark{a} & 7.0 \\   
1.6787        & \ion{Fe}{2}   & -0.046\tablenotemark{a} & 4.2 \\   
1.6802        & Br7           & -0.719\tablenotemark{a} & 66.2 \\
1.6873        & \ion{Fe}{2}   & -0.309 & 28.4 \\   
1.6894        & \ion{Fe}{2}   & -0.015\tablenotemark{a} & 1.4 \\   
1.7006        & \ion{He}{1}   & -0.032\tablenotemark{a} & 2.9 \\   
1.7111        & [\ion{Fe}{2}] & -0.013 & 1.2 \\   
1.7357        & Br6           & -0.521 & 47.1 \\
1.7414        & \ion{Fe}{2}   & -0.102 & 9.2  \\   
1.7451        & [\ion{Fe}{2}] & -0.054\tablenotemark{a} & 4.9 \\   
\hline
1.9440        & Br$\delta$    & -0.808\tablenotemark{b} & 70.3 \\
1.9738        & \ion{Fe}{2}   & -0.019 & 1.7 \\   
2.0590        & \ion{Mg}{2}   & -0.219\tablenotemark{b} & 19.1 \\   
2.1376        & \ion{Mg}{2}   & -0.051 & 4.5 \\   
2.1447        & \ion{Mg}{2}   & -0.018 & 1.6 \\   
2.1649        & Br$\gamma$    & -0.705\tablenotemark{b} & 62.5 \\
2.3532        & Pf23          & -0.028 & 2.5 \\
2.3591        & Pf22          & -0.036 & 3.2 \\
2.3657        & Pf21          & -0.036 & 3.2 \\
2.3731        & Pf20          & -0.054 & 4.9 \\
2.3815        & Pf19          & -0.064 & 5.8 \\
2.3912        & Pf18          & -0.078 & 7.0 \\
2.4023        & Pf17          & -0.075 & 6.7 \\
2.4151        & Pf16          & -0.092 & 8.2 \\
2.4300        & Pf15          & -0.114 & 10.2 \\
2.4477        & Pf14          & -0.162 & 14.4 \\
2.4687        & Pf13          & -0.245 & 21.5 
\enddata
\tablenotetext{a}{These lines suffer from line-blending, which introduces 
  some uncertainty to the derived EW values.}
\tablenotetext{b}{Due to the high $K$-band continuum flux and line brightness, 
  these lines likely reached the non-linearity regime of the FIRE detector,
  resulting in an underestimation of the real EW.}
\end{deluxetable}

Finally, we investigated whether we find indications for variability 
of the {\BrG}-line during the $\sim 6$~months covered by our CRIRES and 
AMBER observations.
For this purpose, we convolved the CRIRES and AMBER HR-K data to the
same spectral resolution and overplot the line profiles in Fig.~\ref{fig:linevariability},
finding no indications for variability, neither in line strength nor line profile.

\subsection{The circumprimary disk}
\label{sec:interpdisk}

Using a temperature-gradient disk model (DISK), we determine the inner continuum disk radius $R_{\rm in}$ to 1.59~mas.
We can compare this value with the expected location of the dust sublimation radius, using the
luminosity and distance estimate of \citet{bor07}, $L_{\star}=(10\pm3) \times 10^{3} L_{\sun}$ 
and $d=1150\pm150$~pc.
Assuming grey dust opacities and a standard dust sublimation temperature of $T_{\mathrm{subl}}=1500$\,K, 
the expected dust sublimation radius (including the effect of backwarming from the disk) is 
$R_{\mathrm{subl}}=\sqrt{L_{\star}/4\pi\sigma T_{\mathrm{subl}}^{4}}$ \citep{dul01}, 
where $\sigma$ denotes the Stefan-Boltzmann constant.
With the given large luminosity and distance 
uncertainties, this corresponds to a value of $6.9\pm1.1$~AU or $6.0\pm1.9$~mas,
which is considerably larger than the measured inner disk radius of $R_{\rm in}=1.59$~mas.
Assuming higher dust sublimation temperatures of 2000\,K 
($R_{\mathrm{subl}}=3.9 \pm 0.6$~AU or $3.4 \pm 1.0$~mas)
also does not solve this discrepancy.
Therefore, we conclude that V921\,Sco belongs to the group of Herbig~Be disks that are
``undersized'' with respect to the size-luminosity relation \citep{mon02},
as already discussed by \citet{kra08b} and \citet{kre12}.
Different physical scenarios have been proposed in order to explain this effect,
including gas absorption \citep{mon02}, the emission from gas located inside 
of the dust sublimation radius \citep{eis04,mon05,kra08a,tan08},
or the presence of a highly refractive dust grain species \citep{ben10}.
Multi-wavelength interferometric studies with even longer baseline lengths
as well as detailed gas and dust radiative transfer modeling will
be required in order to ultimately settle this question.

We determine the outer disk radius to $\sim 9.7$~mas.
However, this value should be treated as a lower limit, since colder 
parts of the disk might extend to larger stellocentric radii.
We can compare this derived lower limit with theoretical 
predictions for the disk truncation radius in binary systems.
\citet{art94} computed the resonance-induced truncation radius
for circumstellar and circumbinary disks and predicts the
circumprimary disk to be truncated at $r/a \sim 40...20$\% of the
orbit major axis (for a binary mass ratio of 0.3), 
depending on the orbit eccentricity and disk viscosity parameter.
This range agrees reasonably well with our derived 
value of $R_{\rm out}/\rho = 39$\%.
For a detailed comparison, a full orbit determination and
mid-infrared and sub-millimeter interferometric observations
of the colder material in the outer regions of the circumprimary disk 
and in the circumbinary gas and dust reservoir will be required.

Both theoretical \citep[e.g.][]{nat01,dul01,ise05} and observational studies \citep{mon06,kra09b,ben11} suggest that
the inner dust rim in protoplanetary disks has a complicated, possibly puffed-up vertical structure,
which can result in an asymmetric brightness distribution.
Interferometric data, such as presented in this study, is able to detect these asymmetries and to 
constrain the corresponding disk models, in particular using phase-closure capabilities.
However, it is clear that the closure phases in our AMBER data are dominated by the signatures
of the newly-detected companion, while potential asymmetries due to the vertical disk structure would 
 appear only as secondary effects.
Such signatures would appear in particular at long baseline lengths and close to visibility minima,
where we also observe some significant deviations from our simple flat disk+companion model 
(Fig.~\ref{fig:modelCP}).
However, it would be difficult to constrain these more complex physical models
with the current uncertainties on the stellar parameters and the limited available $uv$-coverage.
Therefore, we leave it to future investigations to better constrain the inner dust disk geometry around V921\,Sco~A.
Alternatively, the remaining residuals might indicate the orbital motion
of the companion within the five months during which the 2008 AMBER LR-HK data has been recorded
(see Tab~1 in Paper~I).

\begin{figure*}
  \centering
  \includegraphics[angle=0,scale=0.8]{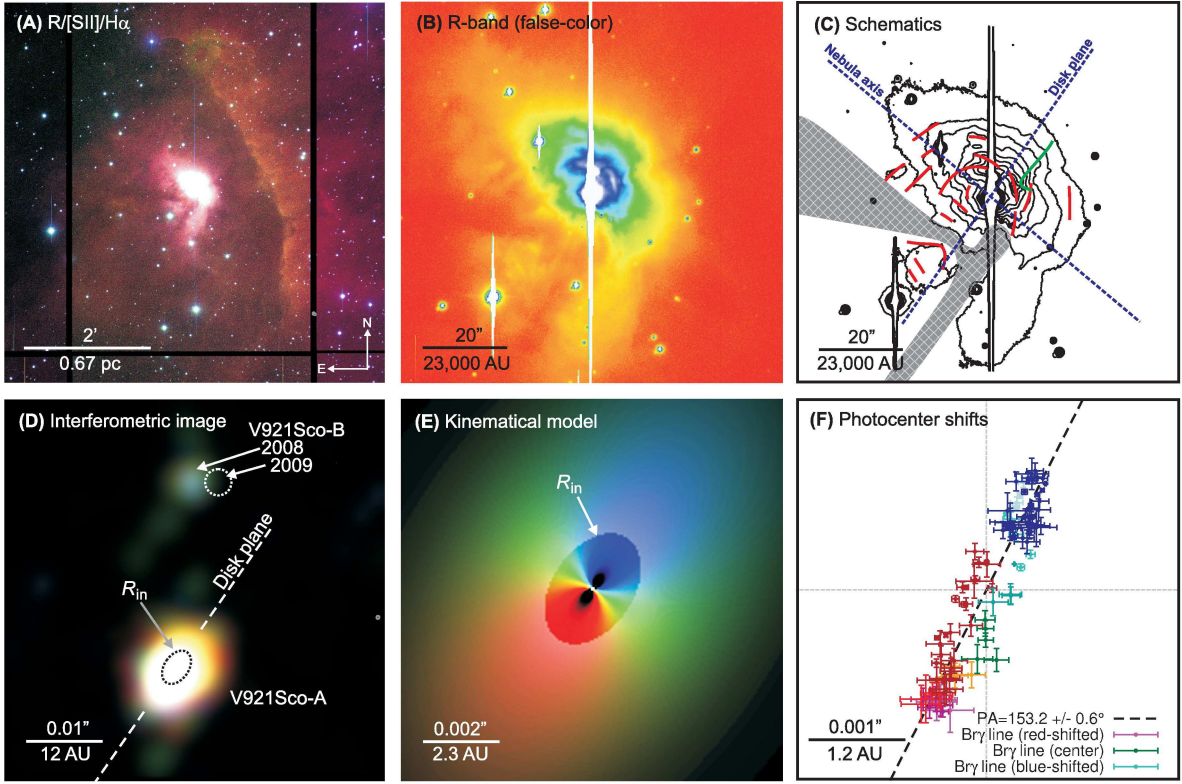}
  \caption{
    Zoom-in on V921\,Sco, covering spatial scales over more than five orders of magnitude.
    On scales of a few arcminutes {\it (A)}, V921\,Sco is embedded in an extended ridge-like structure,
    as shown in this Magellan/IMACS R/{\ForbSII}/{\Ha} image (Paper~I).
    On smaller scales {\it (B)}, V921\,Sco is surrounded by a bipolar nebula with intruiging substructure {\it (C)}.
    The axis of the nebula is perpendicular to the plane of the disk discovered on scales of a few milliarcsecond
    within our interferometric image {\it (D)}.  The black ellipse illustrates the location of the 
    inner continuum disk truncation radius $R_{\rm in}$, as determined with model fitting 
    (Tab.~\ref{tab:modelfitting}, model DISK).
    Most of the {\BrG}-line emission emerges from inside of the dust sublimation radius,
    as determined with our kinematical model {\it (E)}, which is constrained simultaneously 
    by the VLTI/AMBER visibilities, VLTI/AMBER+VLT/CRIRES differential phases, and
    the high-resolution line profile. The image shown here is a color-composite of our model channel maps
    for gas velocities of -35 (blue), 0 (green), and +35~km\,s$^{-1}$ (red).
    The gas rotation is also revealed in a model-independent fashion by the measured photocenter offsets {\it (F)}.
  }
  \label{fig:overview}
\end{figure*}

Using our kinematical modeling, we show that the velocity field in the disk is 
in Keplerian rotation (Sect.~\ref{sec:linemodeldisk}) and that the line-emission 
emerges from a relatively high-density region ($N_e = 2...6\times 10^{19}$\,m$^{-3}$; Sect.~\ref{sec:interpFIRE}),
likely in a hot disk surface layer.
Between the high-level ($u>16$, i.e.\ Pa14-24, Br13-29, and Pf12-23)
and low-level transitions ($u\lesssim 16$, i.e.\ {\PaB}-Pa13, {\BrG}-Br12),
we detect a change of slope in the Paschen/Brackett decrements, 
indicating that the high-level transitions emerge from high-density
regions closer to the star than the low-level transitions.
In our best-fit model (Tab.~\ref{tab:linemodelfitting}, Fig.~\ref{fig:modelKEPLER}), 
the orientation of the opacity screen is such 
that the south-western part of the disk appear brighter, which suggests 
that the north-eastern disk axis is facing towards us
(which is also consistent with the orientation of the large-scale bipolar nebula, Fig.~\ref{fig:overview}, {\it A-C}).
Arguably the most surprising parameter value in our best-fit model
concerns the derived stellar mass.  For an assumed distance of 1.15~kpc, we yield
best agreement with $5.4 \pm 0.4~M_{\sun}$, 
which is hardly consistent with the spectral 
classification\footnote{It should be noted that the earlier spectroscopic
studies were not informed about the binarity of V921\,Sco, 
which has likely also resulted in a somewhat biased 
spectral classifications of this intriguing source.  
However, given the extremely rich recombination line spectrum
and the presence of \ion{He}{1} emission lines, it
seems safe to assume that the ionizing central source
should be of at least B-type.}
and the prediction mass of $9 \pm 1~M_{\sun}$ \citep{bor07}.

In order to explain this surprising result, we propose two plausible scenarios:
\begin{itemize}
\item[(1)] The value of $1.15 \pm 0.15$~kpc \citep{bor07} might
  underestimate the real distance to V921\,Sco. This distance has been determined based 
  on the equivalent width of the \ion{Na}{1} and \ion{Ca}{2}-K interstellar 
  absorption lines and therefore relies on a statistically 
  established calibration. In particular, \citet{lop92} employed the same method
  (also on the \ion{Na}{1} line) and determined the distance to 2.5~kpc.
  A distance estimate $\gtrsim 2$~kpc has also been obtained by \citet{mcg88} 
  based on optical/near-infrared spectroscopy and photometry.
  Adopting such a larger distance would result in a stellar mass 
  of 12~$M_{\sun}$ (for 2.5~kpc) or 9~$M_{\sun}$ (for 2~kpc), 
  which would be more consistent with the spectral classification.

\item[(2)] The disk might not be associated with the massive (early B-type)
  star in the system, but orbit around the intermediate-mass (late B-type) 
  companion instead.
  This would explain the relatively low derived mass, while the massive primary
  star could still provide sufficient ultraviolet flux to ionize the disk
  surface layer and the ambient low-density material resulting in the
  forbidden line-emission.  In this scenario, the accretion would occur
  from the circumbinary disk onto the disk around the late B-type star.
  The massive primary would rest closer to the center-of-mass and be
  effectively shielded from mass-infall by the orbiting companion.
  A similar dynamical configurations seems also plausible for other
  Herbig~B[e]-star binary systems such as MWC\,361~A, where the 
  active accretion is clearly associated with the less massive
  component in the system \citep{mon06,ale08}.
\end{itemize}

In order to distinguish between these scenarios, it will be necessary
to accurately measure the flux ratio of the two stars as function of wavelength.
Our current measurement yield similar flux ratios between the two stars
in the $H$- and $K$-band, with some indications that V921\,Sco~A 
(i.e.\ the star associated with the circumstellar material) is of 
earlier spectral type (Paper~I), favouring scenario (1).  
However, it should be noted that this color measurement 
is only significant at the 2$\sigma$-level
($(F_{\rm B}/F_{\rm A})_{H}=0.83\pm0.15$, 
$(F_{\rm B}/F_{\rm A})_{K}=1.18\pm0.12$, Paper~I) and might be biased, for instance
in case of inhomogeneous extinction towards the two stars.
In both scenarios, it is clear that a careful re-evaluation of
the spectral classification based on the spectral energy distribution
and the spectroscopic diagnostics will be necessary, which,
is out of the scope of this paper.

Independent of these uncertainties, we can
derive new insights on the evolutionary status of V921\,Sco
from our constraints on the disk gas velocity field.
The decretion disks around post-main-sequence supergiant B[e] stars
should exhibit a strong outflowing velocity component \citep{lam91},
which is not observed in our data.
On the other hand, our observation of a Keplerian rotation profile
is consistent with the expected velocity field in a viscous accretion disk \citep{sha73}
around a pre-main-sequence Herbig~B[e] star.

\subsection{On the origin of the {\BrG}-emission in Herbig~Be stars}
\label{sec:BrGorigin}

Arguably, the most important advancement provided by our observational data set on V921\,Sco is
that it allows us to constrain simultaneously the geometry of the circumstellar dust disk
and the spatial distribution and kinematics of the hot gas in the system (see Fig.~\ref{fig:modelKEPLER}).
Such constraints are important, in particular as earlier spatially unresolved studies
have associated the origin of the {\BrG}-line emission with fundamentally different physical mechanisms,
including mass infall (e.g.\ magnetospheric accretion, gaseous inner disks) and mass outflow processes 
(e.g.\ stellar winds, X-winds, or disk winds).
For V921\,Sco, we find that the {\BrG}-emitting gas clearly shows a rotation-dominated velocity field 
and orbits the star in the same plane 
($\varphi=145.0 \pm 1.3^{\circ}$, $i=54 \pm 2^{\circ}$, see Tab.~2)
as the circumstellar dust disk ($\varphi=145.0 \pm 8.4^{\circ}$, $i=48.8\pm 4^{\circ}$), as illustrated in Fig.~\ref{fig:overview}.
On the other hand, \citet{ben10} found for the Herbig~Be-star component in the \object{Z\,CMa} binary system
that the {\BrG}-line emission was displaced along the outflow axis, 
i.e.\ presumably perpendicular to the disk plane.
Given that the Z\,CMa observations were obtained during an exceptionally strong outbursting phase of the
Z\,CMa system in 2008, they likely do not represent typical conditions in a Herbig~Be star.
Another example is provided by \object{MWC\,297}, where \citet{wei11} could reproduce 
important features of the measured wavelength-differential phases using a
magnetospheric disk wind model.
These examples, V921\,Sco (where {\BrG} was found to trace orbiting gas in the disk plane),
Z\,CMa (where the {\BrG}-emitting gas was observed in a bipolar wind),
and MWC\,297 (where the {\BrG}-emitting might trace a disk wind),
illustrate that the {\BrG} emission is probably not associated with a single astrophysical process,
but can trace both mass accretion and outflow processes,
as we have already concluded from our small VLTI/AMBER {\BrG}-survey 
with spectral resolution $R=1500$ \citep{kra08b}.
Spatially and spectrally resolved observations on a larger object sample
will be essential in order to explore under which circumstances the {\BrG}-line
traces a certain mechanism and to determine the relation with spectroscopic diagnostics.

\section{Conclusions}
\label{sec:conclusions}

In this paper, we have investigated the milliarcsecond-scale
environment around the B[e] star V921\,Sco and 
constrained the spatial distribution and kinematics of
ionized hydrogen gas in the system with a spectral resolution up to 
$R=100,000$.  We summarize our findings as follows:

\begin{itemize}
\item Our $H$- and $K$-band continuum interferometric images 
  reveal a spatially extended (Gaussian FWHM $7.5$~mas) disk-like structure, 
  seen under an intermediate inclination angle of $\sim 48.8 \pm 4$\deg.
  Using a temperature-gradient model, we determine the dust sublimation radius to $1.59 \pm 0.25$~mas
  and find that the apparent disk size increases with wavelength, 
  consistent with an irradiated dust disk.

\item Our VLTI/AMBER ($R=12,000$) and VLT/CRIRES ($R=100,000$) observations
  spatially and spectrally resolve the {\BrG}-line emission 
  from V921\,Sco on sub-milliarcsecond scales.
  Using a model-independent photocenter analysis technique and
  our detailed kinematical modeling, we find that the line-emitting
  gas is located in a Keplerian-rotating disk, which extends 
  down to a few stellar radii.
  We interpret this finding as strong evidence for the pre-main-sequence
  nature of the object, since the decretion disks in post-main-sequence 
  B[e]-stars are believed to exhibit a significant outflowing velocity component.

\item From our disk kinematical modeling, we derive a mass of 
  $5.4\pm 0.4~M_{\sun} \cdot (d/1150~\mathrm{pc})$ for the central object.
  Assuming the distance of 1.15~kpc proposed by \citet{bor07}, we find that
  this mass is too low to be consistent with the early B-type spectral classification,
  which might indicate that the distance to V921\,Sco is probably considerably larger ($\sim 2$~kpc).
  Alternatively, the disk might be associated with the less massive component in the system.  
  In this dynamical scenario, the material from the circumbinary disk is accreted 
  onto the intermediate-mass (late B-type) component, preventing significant accretion 
  onto the early B-type star.
  In any case, a careful re-evaluation of the spectral classification
  of the components in this enigmatic binary system will be required in future studies.
  
\item Our FIRE observations reveal a rich near-infrared emission line spectrum
  of permitted and forbidden emission lines (\ion{Fe}{2}, [\ion{Fe}{2}], \ion{C}{1}, \ion{He}{1}, 
  \ion{O}{1}, \ion{N}{1}, \ion{Mg}{2}, \ion{Al}{2}), but show no sign of CO bandhead
  emission, which supports a pre-main-sequence nature \citep{kra09c}.

\item From the FIRE spectra, we derive the line flux for 61 hydrogen recombination lines,
  including transitions from the Paschen, Brackett, and Pfund series.
  Utilizing the spatial information provided by our VLTI/AMBER spectro-interferometric data
  allows us to model the line decrements with a hydrogen excitation model
  and to derive electron densities for the line-emitting gas.
  The derived high number densities ($N_{e}=2...6\times 10^{19}$\,m$^{-3}$)
    suggest that the disk around V921\,Sco is particularly massive,
    reinforcing the conclusion of \citet{hen98}, who derived 
    a total gas mass of $40~M_{\sun}$ for the surrounding millimeter core.
\end{itemize}

Our combined spectro-interferometry \& line decrement modeling approach provides a novel tool
to estimate the gas density in the innermost AU of protoplanetary disks.
Applying this tool on a larger sample of T~Tauri, Herbig~Ae/Be, and transitional disks
might allow, for instance, to quantify gas depletion as function of disk evolution,
or to confirm the expected relations between the mass accretion rate, 
disk surface density, and stellar mass \citep[e.g.][]{cal04}.

Furthermore, our study demonstrates the power of combining infrared
spectro-interferometry with the technique of spectro-astrometry.  
Spectro-interferometry is indispensable to characterize the continuum geometry 
(measurement of the dust disk geometry and,
for instance, the detection of companions) and to resolve the detailed geometry and kinematics of the circumstellar gas.
On the other hand, spectro-astrometry provides a ressource-efficient and straightforward method 
to measure first-order kinematical information with very high spectral dispersion in the spatially unresolved regime, 
providing highly complementary constraints for the interpretation 
of spectro-interferometric signatures.

\acknowledgments
We thank the anonymous referee for constructive comments,
which helped to improve the manuscript.
This work was done in part under contract with the California
Institute of Technology (Caltech), funded by NASA through the 
Sagan Fellowship Program (S.K.\ is a Sagan Fellow).

{\it Facilities:} \facility{VLTI}, \facility{VLT}, \facility{Magellan}.

\bibliographystyle{apj}

\begin{thebibliography}{68}
\expandafter\ifx\csname natexlab\endcsname\relax\def\natexlab#1{#1}\fi

\bibitem[{{Acke} \& {van den Ancker}(2006)}]{ack06}
{Acke}, B., \& {van den Ancker}, M.~E. 2006, \aap, 457, 171

\bibitem[{{Acke} {et~al.}(2005){Acke}, {van den Ancker}, \&
  {Dullemond}}]{ack05}
{Acke}, B., {van den Ancker}, M.~E., \& {Dullemond}, C.~P. 2005, \aap, 436, 209

\bibitem[{{Alecian} {et~al.}(2008){Alecian}, {Catala}, {Wade}, {Donati},
  {Petit}, {Landstreet}, {B{\"o}hm}, {Bouret}, {Bagnulo}, {Folsom}, {Grunhut},
  \& {Silvester}}]{ale08}
{Alecian}, E., {et~al.} 2008, \mnras, 385, 391

\bibitem[{{Artymowicz} \& {Lubow}(1994)}]{art94}
{Artymowicz}, P., \& {Lubow}, S.~H. 1994, \apj, 421, 651

\bibitem[{{Bailey}(1998{\natexlab{a}})}]{bai98b}
{Bailey}, J. 1998{\natexlab{a}}, \mnras, 301, 161

\bibitem[{{Bailey}(1998{\natexlab{b}})}]{bai98a}
{Bailey}, J.~A. 1998{\natexlab{b}}, in Presented at the Society of
  Photo-Optical Instrumentation Engineers (SPIE) Conference, Vol. 3355, Society
  of Photo-Optical Instrumentation Engineers (SPIE) Conference Series, ed.
  {S.~D'Odorico}, 932--939

\bibitem[{{Benedettini} {et~al.}(1998){Benedettini}, {Nisini}, {Giannini},
  {Lorenzetti}, {Tommasi}, {Saraceno}, \& {Smith}}]{ben98}
{Benedettini}, M., {Nisini}, B., {Giannini}, T., {Lorenzetti}, D., {Tommasi},
  E., {Saraceno}, P., \& {Smith}, H.~A. 1998, \aap, 339, 159

\bibitem[{{Benisty} {et~al.}(2010){Benisty}, {Malbet}, {Dougados}, {Natta}, {Le
  Bouquin}, {Massi}, {Bonnefoy}, {Bouvier}, {Chauvin}, {Chesneau}, {Garcia},
  {Grankin}, {Isella}, {Ratzka}, {Tatulli}, {Testi}, {Weigelt}, \&
  {Whelan}}]{ben10}
{Benisty}, M., {et~al.} 2010, \aap, 517, L3+

\bibitem[{{Benisty} {et~al.}(2011){Benisty}, {Renard}, {Natta}, {Berger},
  {Massi}, {Malbet}, {Garcia}, {Isella}, {M{\'e}rand}, {Monin}, {Testi},
  {Thi{\'e}baut}, {Vannier}, \& {Weigelt}}]{ben11}
---. 2011, \aap, 531, A84+

\bibitem[{{Borges Fernandes} {et~al.}(2007){Borges Fernandes}, {Kraus}, {Lorenz
  Martins}, \& {de Ara{\'u}jo}}]{bor07}
{Borges Fernandes}, M., {Kraus}, M., {Lorenz Martins}, S., \& {de Ara{\'u}jo},
  F.~X. 2007, \mnras, 377, 1343

\bibitem[{{Brannigan} {et~al.}(2006){Brannigan}, {Takami}, {Chrysostomou}, \&
  {Bailey}}]{bra06}
{Brannigan}, E., {Takami}, M., {Chrysostomou}, A., \& {Bailey}, J. 2006,
  \mnras, 367, 315

\bibitem[{{Calvet} {et~al.}(2004){Calvet}, {Muzerolle}, {Brice{\~n}o},
  {Hern{\'a}ndez}, {Hartmann}, {Saucedo}, \& {Gordon}}]{cal04}
{Calvet}, N., {Muzerolle}, J., {Brice{\~n}o}, C., {Hern{\'a}ndez}, J.,
  {Hartmann}, L., {Saucedo}, J.~L., \& {Gordon}, K.~D. 2004, \aj, 128, 1294

\bibitem[{{Cardelli} {et~al.}(1989){Cardelli}, {Clayton}, \& {Mathis}}]{car89}
{Cardelli}, J.~A., {Clayton}, G.~C., \& {Mathis}, J.~S. 1989, \apj, 345, 245

\bibitem[{{Chelli} {et~al.}(2009){Chelli}, {Utrera}, \& {Duvert}}]{che09}
{Chelli}, A., {Utrera}, O.~H., \& {Duvert}, G. 2009, \aap, 502, 705

\bibitem[{{Cidale} {et~al.}(2001){Cidale}, {Zorec}, \& {Tringaniello}}]{cid01}
{Cidale}, L., {Zorec}, J., \& {Tringaniello}, L. 2001, \aap, 368, 160

\bibitem[{{Clark} {et~al.}(1999){Clark}, {Steele}, {Fender}, \& {Coe}}]{cla99}
{Clark}, J.~S., {Steele}, I.~A., {Fender}, R.~P., \& {Coe}, M.~J. 1999, \aap,
  348, 888

\bibitem[{{Damineli} {et~al.}(1998){Damineli}, {Stahl}, {Kaufer}, {Wolf},
  {Quast}, \& {Lopes}}]{dam98}
{Damineli}, A., {Stahl}, O., {Kaufer}, A., {Wolf}, B., {Quast}, G., \& {Lopes},
  D.~F. 1998, \aaps, 133, 299

\bibitem[{{de Winter} \& {The}(1990)}]{dew90}
{de Winter}, D., \& {The}, P.~S. 1990, \apss, 166, 99

\bibitem[{{Dullemond} {et~al.}(2001){Dullemond}, {Dominik}, \& {Natta}}]{dul01}
{Dullemond}, C.~P., {Dominik}, C., \& {Natta}, A. 2001, \apj, 560, 957

\bibitem[{{Dullemond} \& {Monnier}(2010)}]{dul10}
{Dullemond}, C.~P., \& {Monnier}, J.~D. 2010, \araa, 48, 205

\bibitem[{{Eisner} {et~al.}(2004){Eisner}, {Lane}, {Hillenbrand}, {Akeson}, \&
  {Sargent}}]{eis04}
{Eisner}, J.~A., {Lane}, B.~F., {Hillenbrand}, L.~A., {Akeson}, R.~L., \&
  {Sargent}, A.~I. 2004, \apj, 613, 1049

\bibitem[{{Gai} {et~al.}(2004){Gai}, {Menardi}, {Cesare}, {Bauvir}, {Bonino},
  {Corcione}, {Dimmler}, {Massone}, {Reynaud}, \& {Wallander}}]{gai04}
{Gai}, M., {et~al.} 2004, in Presented at the Society of Photo-Optical
  Instrumentation Engineers (SPIE) Conference, Vol. 5491, Society of
  Photo-Optical Instrumentation Engineers (SPIE) Conference Series, ed.
  {W.~A.~Traub}, 528--+

\bibitem[{{Goto} {et~al.}(2012){Goto}, {Carmona}, {Linz}, {Stecklum},
  {Henning}, {Meeus}, \& {Usuda}}]{got12}
{Goto}, M., {Carmona}, A., {Linz}, H., {Stecklum}, B., {Henning}, T., {Meeus},
  G., \& {Usuda}, T. 2012, \apj, 748, 6

\bibitem[{{Habart} {et~al.}(2003){Habart}, {Testi}, {Natta}, \&
  {Vanzi}}]{hab03}
{Habart}, E., {Testi}, L., {Natta}, A., \& {Vanzi}, L. 2003, \aap, 400, 575

\bibitem[{{Hase} {et~al.}(2010){Hase}, {Wallace}, {McLeod}, {Harrison}, \&
  {Bernath}}]{has10}
{Hase}, F., {Wallace}, L., {McLeod}, S.~D., {Harrison}, J.~J., \& {Bernath},
  P.~F. 2010, \jqsrt, 111, 521

\bibitem[{{Henning} {et~al.}(1998){Henning}, {Burkert}, {Launhardt}, {Leinert},
  \& {Stecklum}}]{hen98}
{Henning}, T., {Burkert}, A., {Launhardt}, R., {Leinert}, C., \& {Stecklum}, B.
  1998, \aap, 336, 565

\bibitem[{{Hillenbrand} {et~al.}(1992){Hillenbrand}, {Strom}, {Vrba}, \&
  {Keene}}]{hil92}
{Hillenbrand}, L.~A., {Strom}, S.~E., {Vrba}, F.~J., \& {Keene}, J. 1992, \apj,
  397, 613

\bibitem[{{Hutsemekers} \& {van Drom}(1990)}]{hut90}
{Hutsemekers}, D., \& {van Drom}, E. 1990, \aap, 238, 134

\bibitem[{{Isella} \& {Natta}(2005)}]{ise05}
{Isella}, A., \& {Natta}, A. 2005, \aap, 438, 899

\bibitem[{{Jaschek} {et~al.}(1992){Jaschek}, {Jaschek}, {Andrillat}, \&
  {Houziaux}}]{jas92}
{Jaschek}, M., {Jaschek}, C., {Andrillat}, Y., \& {Houziaux}, L. 1992, \mnras,
  254, 413

\bibitem[{{Kaeufl} {et~al.}(2004){Kaeufl}, {Ballester}, {Biereichel},
  {Delabre}, {Donaldson}, {Dorn}, {Fedrigo}, {Finger}, {Fischer}, {Franza},
  {Gojak}, {Huster}, {Jung}, {Lizon}, {Mehrgan}, {Meyer}, {Moorwood}, {Pirard},
  {Paufique}, {Pozna}, {Siebenmorgen}, {Silber}, {Stegmeier}, \&
  {Wegerer}}]{kae04}
{Kaeufl}, H., {et~al.} 2004, in Society of Photo-Optical Instrumentation
  Engineers (SPIE) Conference Series, Vol. 5492, Society of Photo-Optical
  Instrumentation Engineers (SPIE) Conference Series, ed. {A.~F.~M.~Moorwood \&
  M.~Iye}, 1218--1227

\bibitem[{{Kraus}(2009)}]{kra09c}
{Kraus}, M. 2009, \aap, 494, 253

\bibitem[{{Kraus} {et~al.}(2012{\natexlab{a}}){Kraus}, {Calvet}, {Hartmann},
  {Hofmann}, {Kreplin}, {Monnier}, \& {Weigelt}}]{kra12b}
{Kraus}, S., {Calvet}, N., {Hartmann}, L., {Hofmann}, K.-H., {Kreplin}, A.,
  {Monnier}, J.~D., \& {Weigelt}, G. 2012{\natexlab{a}}, \apjl, 746, L2

\bibitem[{{Kraus} {et~al.}(2009{\natexlab{a}}){Kraus}, {Hofmann}, {Malbet},
  {Meilland}, {Natta}, {Schertl}, {Stee}, \& {Weigelt}}]{kra09b}
{Kraus}, S., {Hofmann}, K., {Malbet}, F., {Meilland}, A., {Natta}, A.,
  {Schertl}, D., {Stee}, P., \& {Weigelt}, G. 2009{\natexlab{a}}, \aap, 508,
  787

\bibitem[{{Kraus} {et~al.}(2008{\natexlab{a}}){Kraus}, {Preibisch}, \&
  {Ohnaka}}]{kra08a}
{Kraus}, S., {Preibisch}, T., \& {Ohnaka}, K. 2008{\natexlab{a}}, \apj, 676,
  490

\bibitem[{{Kraus} {et~al.}(2008{\natexlab{b}}){Kraus}, {Hofmann}, {Benisty},
  {Berger}, {Chesneau}, {Isella}, {Malbet}, {Meilland}, {Nardetto}, {Natta},
  {Preibisch}, {Schertl}, {Smith}, {Stee}, {Tatulli}, {Testi}, \&
  {Weigelt}}]{kra08b}
{Kraus}, S., {et~al.} 2008{\natexlab{b}}, \aap, 489, 1157

\bibitem[{{Kraus} {et~al.}(2009{\natexlab{b}}){Kraus}, {Weigelt}, {Balega},
  {Docobo}, {Hofmann}, {Preibisch}, {Schertl}, {Tamazian}, {Driebe}, {Ohnaka},
  {Petrov}, {Sch{\"o}ller}, \& {Smith}}]{kra09a}
---. 2009{\natexlab{b}}, \aap, 497, 195

\bibitem[{{Kraus} {et~al.}(2012{\natexlab{b}}){Kraus}, {Monnier}, {Che},
  {Schaefer}, {Touhami}, {Gies}, {Aufdenberg}, {Baron}, {Thureau}, {ten
  Brummelaar}, {McAlister}, {Turner}, {Sturmann}, \& {Sturmann}}]{kra12a}
---. 2012{\natexlab{b}}, \apj, 744, 19

\bibitem[{{Kreplin} {et~al.}(2012){Kreplin}, {Kraus}, {Hofmann}, {Schertl},
  {Weigelt}, \& {Driebe}}]{kre12}
{Kreplin}, A., {Kraus}, S., {Hofmann}, K.-H., {Schertl}, D., {Weigelt}, G., \&
  {Driebe}, T. 2012, \aap, 537, A103

\bibitem[{{Lachaume}(2003)}]{lac03}
{Lachaume}, R. 2003, \aap, 400, 795

\bibitem[{{Lamers} \& {Pauldrach}(1991)}]{lam91}
{Lamers}, H.~J.~G., \& {Pauldrach}, A.~W.~A. 1991, \aap, 244, L5

\bibitem[{{Lamers} {et~al.}(1998){Lamers}, {Zickgraf}, {de Winter}, {Houziaux},
  \& {Zorec}}]{lam98}
{Lamers}, H.~J.~G.~L.~M., {Zickgraf}, F., {de Winter}, D., {Houziaux}, L., \&
  {Zorec}, J. 1998, \aap, 340, 117

\bibitem[{{Le Bouquin} {et~al.}(2008){Le Bouquin}, {Bauvir}, {Haguenauer},
  {Sch{\"o}ller}, {Rantakyr{\"o}}, \& {Menardi}}]{leb08}
{Le Bouquin}, J., {Bauvir}, B., {Haguenauer}, P., {Sch{\"o}ller}, M.,
  {Rantakyr{\"o}}, F., \& {Menardi}, S. 2008, \aap, 481, 553

\bibitem[{{Lopes} {et~al.}(1992){Lopes}, {Damineli Neto}, \& {de Freitas
  Pacheco}}]{lop92}
{Lopes}, D.~F., {Damineli Neto}, A., \& {de Freitas Pacheco}, J.~A. 1992, \aap,
  261, 482

\bibitem[{{McGregor} {et~al.}(1988){McGregor}, {Hyland}, \& {Hillier}}]{mcg88}
{McGregor}, P.~J., {Hyland}, A.~R., \& {Hillier}, D.~J. 1988, \apj, 324, 1071

\bibitem[{{M{\'e}rand} {et~al.}(2006){M{\'e}rand}, {Bord{\'e}}, \& {Coud{\'e}
  Du Foresto}}]{mer06}
{M{\'e}rand}, A., {Bord{\'e}}, P., \& {Coud{\'e} Du Foresto}, V. 2006, \aap,
  447, 783

\bibitem[{{Monnier} \& {Millan-Gabet}(2002)}]{mon02}
{Monnier}, J.~D., \& {Millan-Gabet}, R. 2002, \apj, 579, 694

\bibitem[{{Monnier} {et~al.}(2005){Monnier}, {Millan-Gabet}, {Billmeier},
  {Akeson}, {Wallace}, {Berger}, {Calvet}, {D'Alessio}, {Danchi}, {Hartmann},
  {Hillenbrand}, {Kuchner}, {Rajagopal}, {Traub}, {Tuthill}, {Boden}, {Booth},
  {Colavita}, {Gathright}, {Hrynevych}, {Le Mignant}, {Ligon}, {Neyman},
  {Swain}, {Thompson}, {Vasisht}, {Wizinowich}, {Beichman}, {Beletic},
  {Creech-Eakman}, {Koresko}, {Sargent}, {Shao}, \& {van Belle}}]{mon05}
{Monnier}, J.~D., {et~al.} 2005, \apj, 624, 832

\bibitem[{{Monnier} {et~al.}(2006){Monnier}, {Berger}, {Millan-Gabet}, {Traub},
  {Schloerb}, {Pedretti}, {Benisty}, {Carleton}, {Haguenauer}, {Kern},
  {Labeye}, {Lacasse}, {Malbet}, {Perraut}, {Pearlman}, \& {Zhao}}]{mon06}
---. 2006, \apj, 647, 444

\bibitem[{{Natta} {et~al.}(1993){Natta}, {Palla}, {Butner}, {Evans}, \&
  {Harvey}}]{nat93}
{Natta}, A., {Palla}, F., {Butner}, H.~M., {Evans}, II, N.~J., \& {Harvey},
  P.~M. 1993, \apj, 406, 674

\bibitem[{{Natta} {et~al.}(2001){Natta}, {Prusti}, {Neri}, {Wooden}, {Grinin},
  \& {Mannings}}]{nat01}
{Natta}, A., {Prusti}, T., {Neri}, R., {Wooden}, D., {Grinin}, V.~P., \&
  {Mannings}, V. 2001, \aap, 371, 186

\bibitem[{{Petrov} {et~al.}(2007){Petrov}, {Malbet}, {Weigelt}, {Antonelli},
  {Beckmann}, {Bresson}, {Chelli}, {Dugu{\'e}}, {Duvert}, {Gennari},
  {Gl{\"u}ck}, {Kern}, {Lagarde}, {Le Coarer}, {Lisi}, {Millour}, {Perraut},
  {Puget}, {Rantakyr{\"o}}, {Robbe-Dubois}, {Roussel}, {Salinari}, {Tatulli},
  {Zins}, {Accardo}, {Acke}, {Agabi}, {Altariba}, {Arezki}, {Aristidi},
  {Baffa}, {Behrend}, {Bl{\"o}cker}, {Bonhomme}, {Busoni}, {Cassaing},
  {Clausse}, {Colin}, {Connot}, {Delboulb{\'e}}, {Domiciano de Souza},
  {Driebe}, {Feautrier}, {Ferruzzi}, {Forveille}, {Fossat}, {Foy},
  {Fraix-Burnet}, {Gallardo}, {Giani}, {Gil}, {Glentzlin}, {Heiden},
  {Heininger}, {Hernandez Utrera}, {Hofmann}, {Kamm}, {Kiekebusch}, {Kraus},
  {Le Contel}, {Le Contel}, {Lesourd}, {Lopez}, {Lopez}, {Magnard}, {Marconi},
  {Mars}, {Martinot-Lagarde}, {Mathias}, {M{\`e}ge}, {Monin}, {Mouillet},
  {Mourard}, {Nussbaum}, {Ohnaka}, {Pacheco}, {Perrier}, {Rabbia}, {Rebattu},
  {Reynaud}, {Richichi}, {Robini}, {Sacchettini}, {Schertl}, {Sch{\"o}ller},
  {Solscheid}, {Spang}, {Stee}, {Stefanini}, {Tallon}, {Tallon-Bosc}, {Tasso},
  {Testi}, {Vakili}, {von der L{\"u}he}, {Valtier}, {Vannier}, \&
  {Ventura}}]{pet07}
{Petrov}, R.~G., {et~al.} 2007, \aap, 464, 1

\bibitem[{{Pi{\'e}tu} {et~al.}(2007){Pi{\'e}tu}, {Dutrey}, \&
  {Guilloteau}}]{pie07}
{Pi{\'e}tu}, V., {Dutrey}, A., \& {Guilloteau}, S. 2007, \aap, 467, 163

\bibitem[{{Pontoppidan} {et~al.}(2011){Pontoppidan}, {Blake}, \&
  {Smette}}]{pon11}
{Pontoppidan}, K.~M., {Blake}, G.~A., \& {Smette}, A. 2011, \apj, 733, 84

\bibitem[{{Shakura} \& {Sunyaev}(1973)}]{sha73}
{Shakura}, N.~I., \& {Sunyaev}, R.~A. 1973, \aap, 24, 337

\bibitem[{{Shore} {et~al.}(1990){Shore}, {Brown}, {Bopp}, {Robinson},
  {Sanduleak}, \& {Feldman}}]{sho90}
{Shore}, S.~N., {Brown}, D.~N., {Bopp}, B.~W., {Robinson}, C.~R., {Sanduleak},
  N., \& {Feldman}, P.~D. 1990, \apjs, 73, 461

\bibitem[{{Simcoe} {et~al.}(2008){Simcoe}, {Burgasser}, {Bernstein}, {Bigelow},
  {Fishner}, {Forrest}, {McMurtry}, {Pipher}, {Schechter}, \& {Smith}}]{sim08}
{Simcoe}, R.~A., {et~al.} 2008, in Society of Photo-Optical Instrumentation
  Engineers (SPIE) Conference Series, Vol. 7014, Society of Photo-Optical
  Instrumentation Engineers (SPIE) Conference Series

\bibitem[{{Smith} \& {Davidson}(2001)}]{smi01}
{Smith}, N., \& {Davidson}, K. 2001, \apjl, 551, L101

\bibitem[{{Storey} \& {Hummer}(1995)}]{sto95}
{Storey}, P.~J., \& {Hummer}, D.~G. 1995, \mnras, 272, 41

\bibitem[{{Takami} {et~al.}(2003){Takami}, {Bailey}, \& {Chrysostomou}}]{tak03}
{Takami}, M., {Bailey}, J., \& {Chrysostomou}, A. 2003, \aap, 397, 675

\bibitem[{{Tannirkulam} {et~al.}(2008){Tannirkulam}, {Monnier}, {Millan-Gabet},
  {Harries}, {Pedretti}, {ten Brummelaar}, {McAlister}, {Turner}, {Sturmann},
  \& {Sturmann}}]{tan08}
{Tannirkulam}, A., {et~al.} 2008, \apjl, 677, L51

\bibitem[{{Tatulli} {et~al.}(2007{\natexlab{a}}){Tatulli}, {Isella}, {Natta},
  {Testi}, {Marconi}, {Malbet}, {Stee}, {Petrov}, {Millour}, {Chelli},
  {Duvert}, {Antonelli}, {Beckmann}, {Bresson}, {Dugu{\'e}}, {Gennari},
  {Gl{\"u}ck}, {Kern}, {Lagarde}, {Le Coarer}, {Lisi}, {Perraut}, {Puget},
  {Rantakyr{\"o}}, {Robbe-Dubois}, {Roussel}, {Weigelt}, {Zins}, {Accardo},
  {Acke}, {Agabi}, {Altariba}, {Arezki}, {Aristidi}, {Baffa}, {Behrend},
  {Bl{\"o}cker}, {Bonhomme}, {Busoni}, {Cassaing}, {Clausse}, {Colin},
  {Connot}, {Delboulb{\'e}}, {Domiciano de Souza}, {Driebe}, {Feautrier},
  {Ferruzzi}, {Forveille}, {Fossat}, {Foy}, {Fraix-Burnet}, {Gallardo},
  {Giani}, {Gil}, {Glentzlin}, {Heiden}, {Heininger}, {Hernandez Utrera},
  {Hofmann}, {Kamm}, {Kiekebusch}, {Kraus}, {Le Contel}, {Le Contel},
  {Lesourd}, {Lopez}, {Lopez}, {Magnard}, {Mars}, {Martinot-Lagarde},
  {Mathias}, {M{\`e}ge}, {Monin}, {Mouillet}, {Mourard}, {Nussbaum}, {Ohnaka},
  {Pacheco}, {Perrier}, {Rabbia}, {Rebattu}, {Reynaud}, {Richichi}, {Robini},
  {Sacchettini}, {Schertl}, {Sch{\"o}ller}, {Solscheid}, {Spang}, {Stefanini},
  {Tallon}, {Tallon-Bosc}, {Tasso}, {Vakili}, {von der L{\"u}he}, {Valtier},
  {Vannier}, \& {Ventura}}]{tat07a}
{Tatulli}, E., {et~al.} 2007{\natexlab{a}}, \aap, 464, 55

\bibitem[{{Tatulli} {et~al.}(2007{\natexlab{b}}){Tatulli}, {Millour}, {Chelli},
  {Duvert}, {Acke}, {Hernandez Utrera}, {Hofmann}, {Kraus}, {Malbet},
  {M{\`e}ge}, {Petrov}, {Vannier}, {Zins}, {Antonelli}, {Beckmann}, {Bresson},
  {Dugu{\'e}}, {Gennari}, {Gl{\"u}ck}, {Kern}, {Lagarde}, {Le Coarer}, {Lisi},
  {Perraut}, {Puget}, {Rantakyr{\"o}}, {Robbe-Dubois}, {Roussel}, {Weigelt},
  {Accardo}, {Agabi}, {Altariba}, {Arezki}, {Aristidi}, {Baffa}, {Behrend},
  {Bl{\"o}cker}, {Bonhomme}, {Busoni}, {Cassaing}, {Clausse}, {Colin},
  {Connot}, {Delboulb{\'e}}, {Domiciano de Souza}, {Driebe}, {Feautrier},
  {Ferruzzi}, {Forveille}, {Fossat}, {Foy}, {Fraix-Burnet}, {Gallardo},
  {Giani}, {Gil}, {Glentzlin}, {Heiden}, {Heininger}, {Kamm}, {Kiekebusch}, {Le
  Contel}, {Le Contel}, {Lesourd}, {Lopez}, {Lopez}, {Magnard}, {Marconi},
  {Mars}, {Martinot-Lagarde}, {Mathias}, {Monin}, {Mouillet}, {Mourard},
  {Nussbaum}, {Ohnaka}, {Pacheco}, {Perrier}, {Rabbia}, {Rebattu}, {Reynaud},
  {Richichi}, {Robini}, {Sacchettini}, {Schertl}, {Sch{\"o}ller}, {Solscheid},
  {Spang}, {Stee}, {Stefanini}, {Tallon}, {Tallon-Bosc}, {Tasso}, {Testi},
  {Vakili}, {von der L{\"u}he}, {Valtier}, \& {Ventura}}]{tat07b}
---. 2007{\natexlab{b}}, \aap, 464, 29

\bibitem[{{The} {et~al.}(1994){The}, {de Winter}, \& {Perez}}]{the94}
{The}, P.~S., {de Winter}, D., \& {Perez}, M.~R. 1994, \aaps, 104, 315

\bibitem[{{Thi} {et~al.}(2011){Thi}, {Woitke}, \& {Kamp}}]{thi11}
{Thi}, W., {Woitke}, P., \& {Kamp}, I. 2011, \mnras, 412, 711

\bibitem[{{{\v S}tefl} {et~al.}(2009){{\v S}tefl}, {Rivinius}, {Carciofi}, {Le
  Bouquin}, {Baade}, {Bjorkman}, {Hesselbach}, {Hummel}, {Okazaki}, {Pollmann},
  {Rantakyr{\"o}}, \& {Wisniewski}}]{ste09}
{{\v S}tefl}, S., {et~al.} 2009, \aap, 504, 929

\bibitem[{{Weigelt} {et~al.}(2011){Weigelt}, {Grinin}, {Groh}, {Hofmann},
  {Kraus}, {Miroshnichenko}, {Schertl}, {Tambovtseva}, {Benisty}, {Driebe},
  {Lagarde}, {Malbet}, {Meilland}, {Petrov}, \& {Tatulli}}]{wei11}
{Weigelt}, G., {et~al.} 2011, \aap, 527, A103

\bibitem[{{Whelan} {et~al.}(2005){Whelan}, {Ray}, {Bacciotti}, {Natta},
  {Testi}, \& {Randich}}]{whe05}
{Whelan}, E.~T., {Ray}, T.~P., {Bacciotti}, F., {Natta}, A., {Testi}, L., \&
  {Randich}, S. 2005, \nat, 435, 652

\end{thebibliography}

\end{document}